\documentclass[10pt,a4paper,twocolumn,english,prb,aps,showpacs,floatfix,groupedaddress,superscriptaddress]{revtex4-2}
\usepackage{graphicx,epsfig}
\usepackage{times}
\usepackage{graphics,dcolumn,bm}
\usepackage{amssymb,amsmath,rotate,color}

\usepackage[breaklinks,colorlinks=true,urlcolor=blue,citecolor=blue,linkcolor=blue]{hyperref}
\usepackage{soul}
\usepackage[normalem]{ulem} 
\usepackage{dsfont}

\begin{document}
\newcommand{\be}{\begin{equation}}
\newcommand{\ee}{\end{equation}}
\newcommand{\bearr}{\begin{eqnarray}}
\newcommand{\eearr}{\end{eqnarray}}
\newcommand{\nn}{\nonumber}
\newcommand{\dagg}{{\dagger}}
\newcommand{\vpdag}{{\vphantom{\dagger}}}
\newcommand{\bs}{\boldsymbol}
\newcommand{\up}{\uparrow}
\newcommand{\down}{\downarrow}
\newcommand{\fns}{\footnotesize}
\newcommand{\ns}{\normalsize}
\newcommand{\cdag}{c^{\dagger}}
\newcommand{\N}{_\text{N}}
\newcommand{\h}{_\text{H}}
\newcommand{\mg}{\Gamma_{\!\rm MG}}
\newcommand{\ctg}{\Gamma_{\!\rm CTG}}

\definecolor{red}{rgb}{1.0,0.0,0.0}
\definecolor{green}{rgb}{0.0,1.0,0.0}
\definecolor{blue}{rgb}{0.0,0.0,1.0}

\title{Simplified approach to the magnetic blue shift of Mott gaps}

\author{Mohsen Hafez-Torbati}
\email{mohsen.hafez@tu-dortmund.de}
\affiliation{Condensed Matter Theory, Department of Physics, TU Dortmund University,
44221 Dortmund, Germany}
\author{Frithjof B. Anders}
\email{frithjof.anders@tu-dortmund.de}
\affiliation{Condensed Matter Theory, Department of Physics, TU Dortmund University,
44221 Dortmund, Germany}
\author{G\"otz S. Uhrig}
\email{goetz.uhrig@tu-dortmund.de}
\affiliation{Condensed Matter Theory, Department of Physics, TU Dortmund University,
44221 Dortmund, Germany}

\date{\today}%

\begin{abstract}
The antiferromagnetic ordering in Mott insulators upon lowering the temperature is accompanied by 
a transfer of the single-particle spectral weight to lower energies and a shift of the Mott gap to higher energies (magnetic 
blue shift, MBS).
The MBS is governed by the double exchange and the exchange mechanisms. Both mechanisms enhance the MBS upon increasing the number 
of orbitals. By performing a polynomial fit to numerical dynamical mean-field theory data 
we provide an expansion for the MBS in terms of hopping and
exchange coupling of a prototype Hubbard-Kondo-Heisenberg model and discuss how the results can be generalized for application to realistic Mott or 
charge-transfer insulator materials. This allows estimating the MBS of the charge gap in real materials in an 
extremely simple way avoiding extensive theoretical calculations. The approach is exemplarily applied 
to $\alpha$-MnTe, NiO, and BiFeO$_3$ and an MBS of about $130$ meV, $360$ meV, and $157$ meV is found, respectively.
The values are compared with the previous theoretical calculations and the available experimental data.
Our ready-to-use formula for the MBS simplifies the future studies searching for materials with a strong coupling 
between the antiferromagnetic ordering and the charge excitations, which is paramount to realize a coupled 
spin-charge coherent dynamics at a femtosecond time scale.

\end{abstract}

\maketitle

\section{Introduction}
\label{sec:intro}

Spintronics utilizes the charge and the spin degrees of freedom of electrons for data storage and 
information processing. While it originally relies 
on ferromagnetic bits, a rapidly growing interest for developing a spintronic 
technology based on antiferromagnetic materials has been witnessed in the last decade.
In contrast to their ferromagnetic counterparts, antiferromagnetic materials combine multiple unique properties 
which make them ideal candidates for the construction of the next generation
spintronic devices. Antiferromagnets are resilient to disturbing magnetic fields which supports long-term data retention,
create no stray field allowing high-density memory integration, and display 
a spin dynamics up to 1000 times faster than ferromagnets providing access to terahertz writing speed regime. 
But the faster dynamics due to substantially larger exchange interactions also represents an obstacle for efficient 
probing and manipulating the magnetic states \cite{Kirilyuk2010,Baltz2018,Nemec2018}. 

In recent years, the interaction of electromagnetic 
radiation with antiferromagnetic states has been demonstrated as a powerful tool to detect and to control 
the magnetic order in antiferromagnetic materials \cite{Nemec2018}. A coherent dynamics of the local magnetization in
antiferromagnetic insulators is induced and manipulated by ultrashort laser pulses \cite{Satoh2010,Kampfrath2011,Bossini2014} 
reaching frequencies as high as 20 THz \cite{Bossini2016,Hashimoto2018,Bossini2019}. 
However, this concerns the magnetic properties of the material only. An important step for the further 
development of this research area is the utilization of spin and charge coupling. 
Aiming at an efficient integration of magnonics, spintronics,
and conventional electronics the conversion of magnetic signals
to charge signals and vice-versa \cite{Gillmeister2020,Han2018} is crucial. This makes
ultrafast spin-charge coupling a cornerstone of future information processing.
An essential first step in this direction is to identify materials displaying a strong coupling between the spin and 
the charge degrees of freedom which does not rely on the typically weak spin-orbit coupling. 
One notes that the strength of the spin-charge coupling determines the characteristic time scale at which the 
conversion of spin signals into charge responses can take place and govern the spin-charge oscillation dynamics. 

Mott insulators, or the closely related charge-transfer insulators, are among the promising materials for
antiferromagnetic spintronics \cite{Baltz2018}. In Mott insulators in transition metal compounds strong repulsive interaction between 
electrons in the $3d$ shell is reponsible for the energy gap opening in the single-particle 
spectral function (density of states). This gap defines the energy necessary to create an electron and a hole independent 
from each other in the system and is known as the charge gap or the Mott gap. 
Depending on the number of unpaired electrons in the $3d$ shell one may realize Mott insulators with different number of relevant orbitals. 
Most of the Mott insulators undergo a transition 
from a paramagnetic state to an antiferromagnetic state upon reducing the temperature below a critical value called N\'eel 
temperature $T_{\rm N}$. The magnetic shift of the Mott gap is given by
\be
\label{eq:mbs}
\Gamma_{\rm MG}(T)=G_{\rm MI}(T)-G_{\rm PI}(T),
\ee
where $G_{\rm MI}(T)$ stands for the gap in the magnetic insulator (MI), and $G_{\rm PI}(T)$ stands for
the gap in the paramagnetic insulator (PI) at the temperature $T$. 
A finite value of $\Gamma_{\rm MG}(T)$ represents a direct coupling between the spin and the charge degrees of freedom.
For $T<T_{\rm N}$ the PI is not the stable phase due to the spin ordering but $G_{\rm PI}(T)$ can still be analyzed 
by extrapolating the results to below $T_{\rm N}$ in experiment or by enforcing a paramagnetic solution in theory. 
A magnetic blue shift (MBS) corresponding to $\Gamma_{\rm MG}(T)>0$ is observed in several previous experimental 
\cite{Busch1966,Blazey1969,Wachter1970,Chou1974,Ferrer-Roca2000,Bossini2020,Diouri1985}
and theoretical investigations \cite{Diouri1985,Alexander1976,Ando1992,Sangiovanni2006,Wang2009,Fratino2017,Hafez-Torbati2021} 
of antiferromagnetic insulators. We unveiled in Ref.\ \cite{Hafez-Torbati2021} that the MBS originates from the double exchange 
and exchange mechanisms.
A strong MBS enables to go beyond the local magnetization dynamics \cite{Bossini2016,Hashimoto2018,Bossini2019} 
and induce an ultrafast coherent manipulation of the transport properties.
This highlights the importance of a systematic analysis of the influential mechanisms involved in the MBS of the 
Mott gap \eqref{eq:mbs}. 

This paper is devoted to a systematic study of the effect of antiferromagnetic ordering on 
the single-particle spectral function in Mott insulators. We specifically address how such a spin-charge 
coupling changes upon changing the number of orbitals. 
Our investigation relies on a dynamical mean-field theory (DMFT) 
analysis of a generic three-dimensional Hubbard-Kondo-Heisenberg (HKH) model. 
This model combines a single-band Hubbard model with localized spins $S$ as in a Kondo lattice
model. The latter also represent electrons in orbitals so that we achieve an approximate description 
of a multi-orbital model with $2S+1$ orbitals in total.
The antiferromagnetic ordering is accompanied by a transfer of spectral weight to lower energies 
and a MBS of the Mott gap. The double exchange and exchange mechanisms both enhance the MBS upon increasing the number of orbitals.

For our prototype model we provide an expansion for the MBS in terms of hopping and exchange coupling 
by performing a polynomial fit to numerical DMFT data. We discuss how the expansion can be generalized for 
application to real materials. This allows estimating the MBS of the charge gap in 
Mott as well as in charge-transfer insulator materials in an extremely simple way which avoids extensive theoretical calculations. 
We exemplarily apply the approach to $\alpha$-MnTe, NiO, and BiFeO$_3$ as promising candidates for antiferromagnetic spintronics. 
For $\alpha$-MnTe we find an MBS of about $130$ meV which is in a very good agreement with the previous theoretical calculations 
and experimental data. For NiO and BiFeO$_3$ we obtain an MBS of about $360$ and $157$ meV, respectively, which are also
compared with the available experimental data.
Our ready-to-use formula for the MBS simplifies future investigations searching for materials with strong spin-charge 
coupling which would then set the stage to realize a coupled spin-charge coherent dynamics at the unprecedented 
high frequencies in the THz region.

The paper is organized as follows. In Section \ref{sec:mm} we discuss the model and the method. 
The results are presented in Section \ref{sec:re}. Section \ref{sec:app} is devoted to the application 
of our results to real materials.
The paper is concluded in Section \ref{sec:out}.

\section{Model and method}
\label{sec:mm}

\subsection{Hubbard-Kondo-Heisenberg model}
\label{sec:hkh}

A full description of the charge and the spin degrees of freedom in strongly correlated multi-orbital systems 
requires an analysis of a multi-orbital Hubbard model whose reliable investigation is a grand challenge 
for current theoretical research. Some simplifications can be applied in the case that the system 
is deep in the Mott regime. In Mott insulators the charge excitations are high in energy 
and the low-energy properties of the system are governed by spin excitations. A common and successful 
strategy to address the low-energy excitations is by mapping the multi-orbital Hubbard model
to the Heisenberg model 
\be
\label{eq:hem}
H=J\sum_{\langle i,j \rangle} \vec{\mathcal{S}}_i \cdot \vec{\mathcal{S}}_j
\ee
with the maximum local quantum spin number $\mathcal{S}$ in accord with Hund's first rule. 
The notation $\langle i,j \rangle$ limits the exchange spin-spin interaction to nearest neighbor sites.
However, such a mapping completely removes the charge degrees of freedom and prevents any access to the charge 
excitations which are the main focus of this paper. 

\begin{figure}[t]
   \begin{center}
   \includegraphics[width=0.97\columnwidth,angle=0]{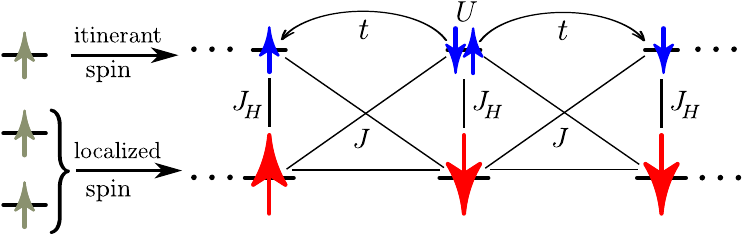}
   \caption{Description of a three-orbital Mott insulator as a Hubbard-Kondo-Heisenberg model 
  \eqref{eq:hkh} with a localized spin $S=1$. The itinerant electron is described by the Hubbard model
  with nearest-neighbor hopping $t$ and onsite repulsion $U>0$. The spin of the itinerant 
  electron is coupled by the Hund coupling $J_{\rm H}$ to the localized spin. There are nearest-neighbor 
  exchange interactions $J$ between electron and localized spins resulting from the exchange processes.}
   \label{fig:hkh}
   \end{center}
\end{figure}

Hence, we have to go beyond the Heisenberg model \eqref{eq:hem} and address the charge gap in multi-orbital Mott insulators
within a tractable model. We follow the idea initialy proposed in Ref.\ \cite{Bossini2020} and applied 
successfully to $\alpha$-MnTe in Ref.\ \cite{Hafez-Torbati2021}. It is discussed here in details for the sake of completeness.
The idea relies on considering one orbital as itinerant and the other orbitals as localized. 
The electron in the itinerant orbital feels the existance of the localized electrons 
via an effective spin-spin interaction, see Fig.\ \ref{fig:hkh}. We stress that the itinerant orbital is a representative for all orbitals. 
It does not mean we make a distinction between the orbitals. 
We will discuss below that this approach takes into account on-site charge fluctuations $\pm1$ around the half filling, but no
further ones as, e.g., $\pm 2$. The system is described by a HKH model given by 
\be
\label{eq:hkh}
H=H_{\rm Hu}+H_{\rm Ko}+H_{\rm He}
\ee
with
\begin{subequations}
 \begin{align}
 \label{eq:hu}
  H_{\rm Hu}=&-t \sum_{\langle i,j\rangle}\sum_{\sigma}( c^{\dagg}_{i,\sigma}c^{\vpdag}_{j,\sigma}+{\rm H.c.}) 
\!+\! U \sum_{i} n^{\vpdag}_{i,\downarrow} n^{\vpdag}_{i,\uparrow} \ , \\
\label{eq:ko}
H_{\rm Ko}=&- 2J\h \sum_{i} \vec{S}_i \cdot \vec{s}_i \ , \\
\label{eq:he}
  H_{\rm He}=& +J\sum_{\langle i,j \rangle} ( \vec{S}_i \cdot \vec{s}_j +
\vec{S}_j \cdot \vec{s}_i +\vec{S}_i \cdot \vec{S}_j) \ ,
 \end{align}
\end{subequations}
where $c^{\dagg}_{i,\sigma}$ and $c^{\vpdag}_{i,\sigma}$ are the usual fermionic creation and annihilation operators 
and $n^{\vpdag}_{i,\sigma}=c^{\dagg}_{i,\sigma}c^{\vpdag}_{i,\sigma}$ is the occupation number operator at the lattice
site $i$ with spin $\sigma$.
The Hubbard model \eqref{eq:hu} with the nearest-neighbor hopping $t$ and the onsite interaction 
$U$ between electrons with opposite spins $\sigma=\uparrow, \downarrow$ describes the itinerant electrons, the Kondo
term \eqref{eq:ko} describes the ferromagnetic Hund coupling $J_{\rm H}>0$ between the electron spin $\vec{s}_i$ and the localized 
spin $\vec{S}_i$ at each lattice site $i$, and the Heisenberg term \eqref{eq:he} describes the virtual exchange
interactions between the itinerant and the localized spins at neighboring sites. The Hamiltonian is schematically 
depicted in Fig.\ \ref{fig:hkh}. The quantum spin number of the localized spin $S$ depends on the number
of orbitals. 
The spin $S$ represents $2S$ orbitals so that including the itinerant orbital $2S+1$ orbitals are described
in total. Hence, we have $S=0$ for the single orbital case and $S=2$ for the five-orbital 
case, which is the maximum value in transition metal compounds.
In order to assess the influence of the number of orbitals we vary $S$. 
For the case $S=0$ there are already data available in the literature. Hence, we mainly focus 
on the representative values $S=1$ and $S=2$ leaving other values to interpolation,
although we present data in Section \ref{sec:he} also for the $S=1/2$ and the $S=3/2$ cases.

Throughout this paper we fix the exchange interaction $J$ in \eqref{eq:he} to $J=4t^2/\Delta$ where $\Delta=U+2SJ_{\rm H}$ is 
the bare charge gap, i.e., the charge gap in the absence of any hopping. This is to guarantee that the low-energy properties 
of Hamiltonian \eqref{eq:hkh} for $U,J_{\rm H}\gg t$ match those of the Heisenberg model \eqref{eq:hem}
with the spin quantum number $\mathcal{S}=S+1/2$ and the exchange interaction 
$J=4t^2/\Delta$.
The exchange interaction $J= 4t^2/\Delta$ is independent from the number of orbitals; 
it describes the exchange coupling between a pair of electrons on linked sites. 
We assume equal intra- and inter-orbital hopping elements and the number of 
contributing electrons is taken into account by the size of the local spin $S$ consistent with 
the low-energy Heisenberg model Eq.\ \eqref{eq:hem}. 
The effective interaction \eqref{eq:he} is limited to the leading 
contribution as the system is considered to be deep in the Mott regime.

The Hamiltonian \eqref{eq:hkh} goes beyond the low-energy Heisenberg model \eqref{eq:hem} by providing 
an appropriate description of the multi-orbital Mott insulators in the one hole and the one double-occupancy subspaces, 
counted relative to half-filling of the orbitals. 
The states which contain more than one hole and one double occupancy at the same lattice site are neglected in the 
Hamiltonian \eqref{eq:hkh}. However, in the limit of large $U$ and $J_{\rm H}$ these states are high in energy 
and are not expected to contribute to the low-lying charge excitations. In addition, we assume that the Hund coupling 
is large enough to ensure that the spins of the local electrons are aligned parallely.
In Section \ref{sec:neel} we compare the low-energy properties of the HKH model \eqref{eq:hkh} and the 
corresponding Heisenberg model \eqref{eq:hem} and discuss the effect of the neglected spin states on the 
thermal fluctuations.

We point out that in general multi-orbital systems the crystal field splitting modifies 
the bare charge gap $\Delta=U+2SJ_{\rm H}$. However, this contribution is negligible compared to 
the intra-orbital interaction $U$ and the Hund coupling $J_{\rm H}$. 
In addition, we only aim at an estimate of the magnetic blues shift deep in the Mott regime at half filling.

It is already an established and efficient method to determine the exchange interactions in Mott insulators 
by fitting theoretical magnon \cite{Coldea2001,Szuszkiewicz2006,Powalski2018} or 
triplon \cite{Knetter2000a,Knetter2004,knett01a,Notbohm2007} dispersions computed for Heisenberg models to 
experimental data, mostly
inelastic neutron scattering data.
The Hubbard interaction and the Hund coupling can be estimated based on atomic physics or density-functional 
theory analysis, which in combination with the exchange interactions allow us to determine the hopping 
via $J=4t^2/\Delta$. Hence, the HKH model \eqref{eq:hkh} provide a well-justified way to estimate
the charge gap in multi-orbital Mott insulators. 
We will estimate hopping elements from the exchange values $J$. The idea is to use the minimum of 
experimental input for the intended estimate. Of course, density functional theory input could also be used. 
But this would require additional demanding calculations.

One should note that different bare charge gaps $\Delta=U+2SJ_{\rm H}$ (and exchange interactions $J=4t^2/\Delta$) 
will be obtained if the same Hubbard interaction $U$ and the same Hund coupling $J_{\rm H}$ are used for 
different spin lengths $S$. 
This prevents a direct comparison of the results obtained for different spin lengths since it is the hopping $t$ 
and the bare charge $\Delta$ (or $t$ and $J=4t^2/\Delta$) which determine the MBS.
For this reason we compare results obtained for $J_{\rm H}=0.3U/S$ and a $U$ independent from $S$. 
In this way, the bare charge gap $\Delta=1.6U$ depends only on $U$ and for constant $U$ the exchange interaction 
stays constant.
The results for the MBS will reflect the influence of the spin length only.
However, in Section \ref{sec:he} we will consider also other combinations of the Hund coupling $J_{\rm H}$ and 
the Hubbard interaction $U$ in order to show that the 
MBS in the Mott regime does not depend on $U$ and $J_{\rm H}$ individually, 
but only on their combination through the bare charge gap $\Delta$.

\subsection{Dynamical mean-field theory}
\label{sec:dmft}

We study the charge gap in both the MI and the PI phases 
of the HKH model on the generic three-dimensional cubic lattice using the DMFT technique. 
The chosen theoretical approach of DMFT is an established method for strongly correlated systems with large 
coordination numbers which fully takes into account the local quantum fluctuations \cite{Georges1996}. 
The self-energy is approximated as local but it
can depend on frequency in contrast to static mean-field theories. We emphasize that a frequency-dependent self-energy 
is essential \cite{Gebhard1997} to describe the paramagnetic Mott insulator and to address the MBS in Eq.\ \eqref{eq:mbs}.
The DMFT is already employed to address Kondo lattice systems \cite{Held2000,Furukawa1994,Yunoki1998}. 
We applied the method specifically to the HKH model in Ref.\ \cite{Hafez-Torbati2021}.
Here, we review the procedure for the sake of completeness.

We assume collinear antiferromagnetic  N\'eel order with the spin polarization in the $z$ direction.
In the limit of large coordination number justifying the DMFT the effect of the intersite 
exchange interactions \eqref{eq:he} can be captured by a mean-field approximation \cite{Hartmann1989} as
\be
\label{eq:he:mf}
H_{\rm He}=-\sum_i h_{i}^{\rm iti} s_i^z -\sum_i h_{i}^{\rm loc} S_i^z
\ee
with the effective magnetic fields $h_{i}^{\rm iti}$ acting on the itinerant 
spin and $h_{i}^{\rm loc}$ acting on the localized spin given by 
\begin{subequations}
 \begin{align}
 \label{eq:hi:iti}
  h_{i}^{\rm iti}&=6J\langle S_i^z \rangle \ , \\
 \label{eq:hi:loc}
  h_{i}^{\rm loc}&=6J\langle S_i^z +s_i^z \rangle \ .
 \end{align}
\end{subequations}
The approximation \eqref{eq:he:mf} reduces the Hamiltonian \eqref{eq:hkh} to a model involving only 
local interactions in addition to electron hopping. 
We expect the mean-field treatment \eqref{eq:he:mf} of the exchange interaction \eqref{eq:he} 
to be more accurate for larger localized spins, where the fluctuations become less relevant.

Although the DMFT formulated in momentum space is suitable for the current problem we opt for the real-space 
DMFT \cite{Potthoff1999,Song2008,Snoek2008} due to its easy adaptability to study the 
spatial dependence of the MBS in thin films in the future as well as due to the efficient and flexible implementation 
we have for the method at hand \cite{Hafez-Torbati2018} which has been already applied to multiple 
problems \cite{Hafez-Torbati2019,Hafez-Torbati2020,Ebrahimkhas2021,Hafez-Torbati2021}.
The main disadvantage of the real-space DMFT is the finiteness of the system. We consider periodic 
$L\times L\times L$ clusters with mainly $L=10$. But, we check that the results even close to the N\'eel 
temperature remain the same with the results obtain for $L=20$. We stress that although we use the 
real-space DMFT we fully exploit the lattice symmetry 
to compute the lattice Green's function and to address the local impurity problem \cite{Hafez-Torbati2018}.
This is the reason we can easily reach large system sizes.

The DMFT requires the solution of a self-consistency problem. Its iteration
loops start with an initial guess for the local polarizations $\langle S_i^z \rangle$ and 
$\langle s_i^z \rangle$ and the local self-energy $\Sigma_{i,\sigma}(i\omega_n)$ at a single representative 
lattice site $i$, which allows us to determine the effective magnetic fields in Eq.\ \eqref{eq:he:mf} and the self-energy
on the whole lattice because the N\'eel phase is invariant under the combination of the spin-up and spin-down 
sublattices swap and the spin-flip transformation.
The lattice Green's function is calculated using the lattice Dyson equation
\be
\label{eq:lgf}
{\bs G}(i\omega_n)
= \left( i\omega_n \mathds{1} -{\bs H}_0- {\bs \Sigma}(i\omega_n) \right)^{-1},
\ee
where $\mathds{1}$ is the identity matrix, ${\bs \Sigma}(i\omega_n)$ denotes the self-energy matrix 
with only non-zero diagonal elements $\Sigma_{i,\sigma}(i\omega_n)=[{\bs \Sigma}(i\omega_n)]_{i,\sigma;i,\sigma}$, and 
${\bs H}_0$ is the matrix representation of the operator
\be
H_0 = -\sum_i (h_{i}^{\rm iti} s_i^z +\mu n_i)
-t \sum_{\langle i,j\rangle}\sum_{\sigma}( c^{\dagg}_{i,\sigma}c^{\vpdag}_{j,\sigma}+{\rm H.c.})
\ee
in the one-particle subspace $\{i,\sigma\}$. 
The chemical potential $\mu=U/2$ is used to satisfy the half-filling condition. One notes that
the lattice Green's function \eqref{eq:lgf} is constructed for itinerant electrons. 

To obtain the local Green's function $\mathcal{G}_{i,\sigma}(i\omega_n)=[{\bs G}(i\omega_n)]_{i,\sigma;i,\sigma}$ we find the two columns of 
the inverse matrix in Eq.\ \eqref{eq:lgf} corresponding to the representative site $i$ with up spin $\sigma=\uparrow$ and 
down spin $\sigma=\downarrow$. The full inversion in Eq.\ \eqref{eq:lgf} is not needed because each column of the inverse 
matrix can be computed independent from the others. The inverse dynamical
Weiss field $\mathcal{G}^{(0)}_{i,\sigma}(i\omega_n)^{-1}$,
which is the local propagator in the auxiliary impurity problem without interaction,
is obtained from the local Dyson equation
\be
\label{eq:dwf}
\mathcal{G}^{(0)}_{i,\sigma}(i\omega_n)^{-1}=\mathcal{G}_{i,\sigma}(i\omega_n)^{-1} +\Sigma_{i,\sigma}(i\omega_n)^{\vphantom{-1}} \ .
\ee
The local impurity problem at the representative site $i$ is described by an effective Anderson-Kondo model 
\bearr
\label{eq:akm}
&&H_i^{\vpdag}=-\mu n^{\vpdag}_i -h_i^{\rm iti} s_i^z 
+ U n^{\vpdag}_{i,\downarrow} n^{\vpdag}_{i,\uparrow}
-h_i^{\rm loc} S_i^z - 2J^\vpdag_{\!H} \vec{S}_i \cdot \vec{s}_i 
\nn \\
&+&\sum_{\ell=1}^{n_b} \sum_{\sigma} \epsilon^{i}_{\ell} 
a^{\dagg}_{\ell,\sigma} a^{\vpdag}_{\ell,\sigma}
+\sum_{\ell=1}^{n_b} \sum_{\sigma} \left(a^{\dagg}_{\ell,\sigma} 
V^i_{\ell,\sigma} c^{\vpdag}_{i,\sigma}+{\rm H.c.} \right) ,
\eearr
where $a^{\dagg}_{\ell,\sigma}$ and $a^{\vpdag}_{\ell,\sigma}$ are the fermionic creation and 
annihilation operators at the bath site $\ell$ with the spin $\sigma=\uparrow,\downarrow$.
The bath sites approximate the effect of the surrounding sites in the lattice.
The bath parameters $\epsilon^{i}_{\ell}$ and $V^i_{\ell,\sigma}$ are determined by
fitting the inverse dynamical Weiss field \eqref{eq:dwf} to the function \cite{Caffarel1994}
\be
\tilde{\mathcal{G}}^{(0)}_{i,\sigma}(i\omega_n)^{-1}=i\omega_n +\mu +{\rm sgn}(\sigma) \frac{h_i^{\rm iti}}{2} 
-\sum_{\ell=1}^{n_b}\frac{|V^{i}_{\ell,\sigma} |^2}{i\omega_n-\epsilon^i_\ell} \ ,
\ee
where ${\rm sgn}(\sigma)=+1$ for $\sigma=\uparrow$ and $-1$ for $\sigma=\downarrow$.
We stress that $\vec{s}_i$ and $\vec{S}_i$ in Eq.\ \eqref{eq:akm} are fully quantum mechanical spin operators.
The local problem \eqref{eq:akm} is solved using exact diagonalization (ED) with a finite number of 
bath sites $n_b$. The local polarizations $\langle S_i^z \rangle$ and $\langle s_i^z \rangle$  
and the local interacting Green's function $\tilde{\mathcal{G}}_{i,\sigma}(i\omega_n)$,
using the Lehmann representation, are computed. The self-energy is updated in each recursive iteration via
\be
\label{eq:sf}
\Sigma_{i,\sigma}(i\omega_n)^{\vphantom{-1}}=\tilde{\mathcal{G}}^{(0)}_{i,\sigma}(i\omega_n)^{-1}-\tilde{\mathcal{G}}_{i,\sigma}(i\omega_n)^{-1} \ ,
\ee
and is used together with the local polarizations for the next iteration. The iterations are continued until 
the convergence within a given tolerance is reached. 

We perform the DMFT iterations mainly with 200 positive Matsubara frequencies, $\omega_n=(2n+1)\pi T$ with $n=0,1, \cdots ,199$. 
However, for the paramagnetic phase at low temperatures we increased the number of positive frequencies even up to 1000. 
This is because in the paramagnetic Mott insulator the self-energy diverges at low temperatures and more 
Matsubara frequencies are needed to achieve accurate results, i.e., to achieve results which no longer depend
on the number of Matsubara frequencies used.

The spin-averaged single-particle spectral function $A(\omega)$ is obtained from the real-frequency Green's function 
of the effective Anderson-Kondo model
\eqref{eq:akm} as
\be
\label{eq:asf}
A(\omega)=-\frac{1}{2}\frac{1}{\pi}\sum_\sigma{\rm Im} \left[ \tilde{\mathcal{G}}_{i,\sigma}(\omega+i\eta) \right]
\ee
where $\eta=0.05t$ is the broadening factor. The spectral function $A(\omega)$ does not depend on the lattice 
site $i$ even in the antiferromagnetic phase since we average over both spin components. 
The ED solver with a finite
number of bath sites approximates a continuous spectral structure by a set of discrete sharp peaks. 
Although the fine details of the spectral function are not produced by the ED solver, the method is 
considered accurate for the charge gap and is used for the single-orbital Hubbard model to benchmark the results 
obtained by quantum Monte Carlo solver \cite{Wang2009} which suffer from the requirement of analytic continution 
to access the real-frequency dynamics. In the case of Anderson-Kondo model the quantum Monte Carlo solver furthermore suffers 
from the notorious fermionic sign problem \cite{Held2000,Peters2006} and is restricted to the Ising part of
the Hund interaction. Therefore, we opt for the ED solver. 

\section{Results}
\label{sec:re}

\subsection{N\'eel temperature}
\label{sec:neel}

Before proceeding to the charge excitations, we first examine the low-energy properties of the HKH model \eqref{eq:hkh}
by computing the N\'eel temperature $T_{\rm N}$ and comparing it with the N\'eel temperature of 
the corresponding Heisenberg model \eqref{eq:hem}.

The DMFT takes into account the local quantum fluctuations but neglects the quantum fluctuations which are non-local. 
The local spin and charge quantum fluctuations in the HKH model \eqref{eq:hkh} are eliminated in the Heisenberg model \eqref{eq:hem}.
The Heisenberg model involves only non-local spin quantum fluctuations.
Thus, we expect a DMFT analysis of the low-energy features of the HKH model \eqref{eq:hkh} to be equivalent to
a mean-field analysis of the corresponding Heisenberg model \eqref{eq:hem}. 
This had already been observed in the single-orbital Hubbard model \cite{Rohringer2018}.

The local mean-field treatment of the Heisenberg model is the same as the mean-field treatment of the corresponding Ising 
model because we consider collinear N\'eel order. 
The mean-field equation for the local spin polarization $m$ of the spin-$\mathcal{S}$ Ising model \cite{Strecka2015} is given by
\be
\label{eq:mft}
m=\mathcal{S}-\frac{\sum_{n=0}^{2\mathcal{S}} n\exp(-ZJmn/T)}{\sum_{n=0}^{2\mathcal{S}}\exp(-ZJmn/T)}
\ee
where $Z$ is the coordination number and $n$ takes integer values. The N\'eel temperature of the spin-$\mathcal{S}$ Ising model in the 
mean-field approximation \cite{Strecka2015} reads
\bearr
\label{eq:tn}
T_{\rm N}=\frac{Z}{3}J\mathcal{S}(\mathcal{S}+1)
=2J(S+1/2)(S+3/2)
\eearr
where the coordination number is given by $Z=6$ for the generic cubic lattice and we substituted $\mathcal{S}=S+1/2$.

\begin{figure}[t]
   \begin{center}
 \includegraphics[width=0.84\columnwidth,angle=-90]{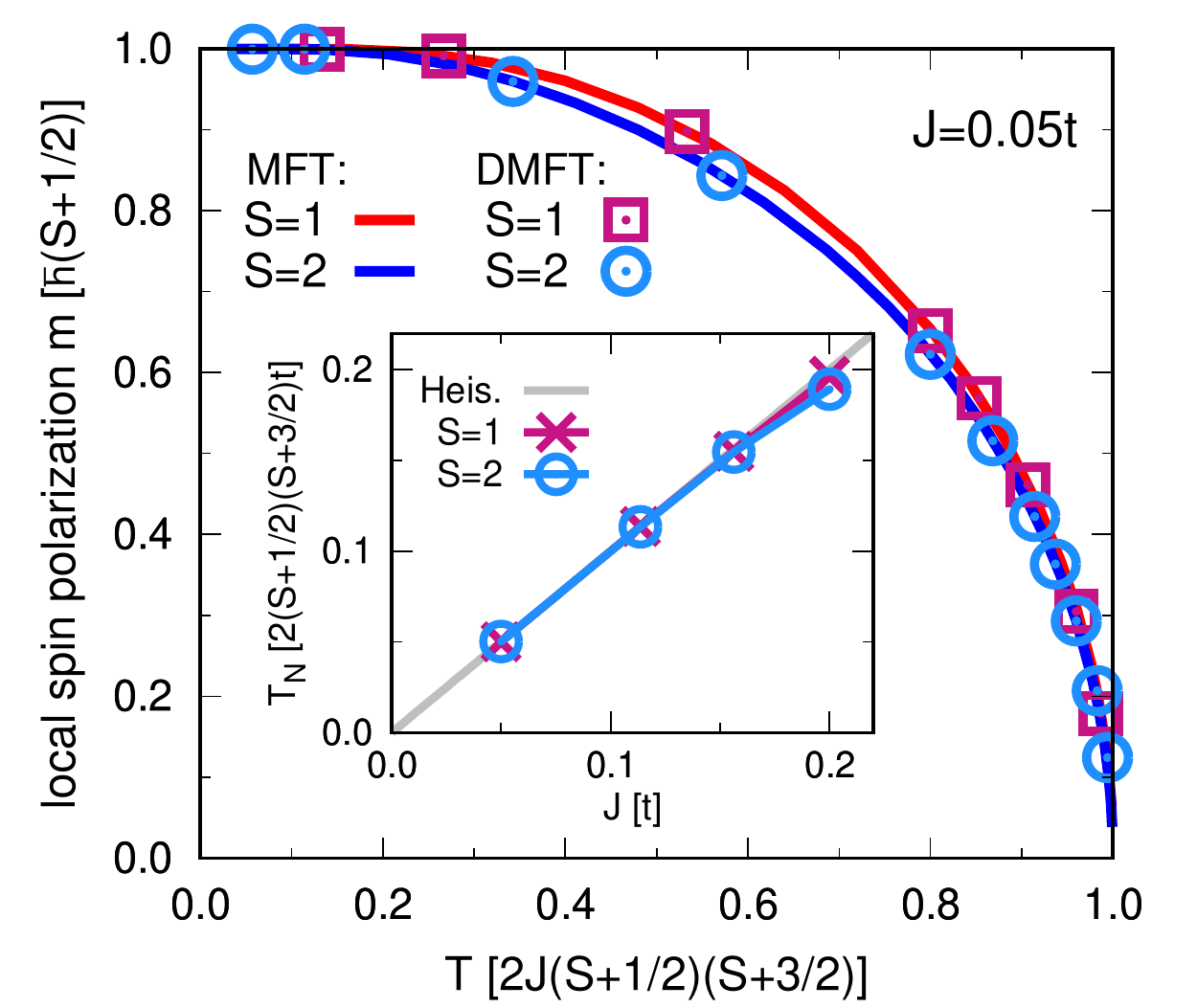}
   \caption{The main plot shows the local spin polarization $m=|\langle S_i^z + s_i^z \rangle|$ vs temperature $T$ 
   in the Hubbard-Kondo-Heisenberg (HKH) model \eqref{eq:hkh} for the localized spins $S=1$ and $S=2$. 
   The DMFT results of the HKH model (symbols) are compared with the mean-field theory (MFT) results
   of the corresponding low-energy Heisenberg model (lines).
   The exchange interaction is fixed to $J=4t^2/\Delta=0.05t$ with the bare charge
   gap $\Delta=U+2SJ_{\rm H}=1.6U$. The inset compares the DMFT N\'eel temperature $T_{\rm N}$ of the HKH model 
   for $S=1$ and $S=2$ with the MFT N\'eel temperature given in Eq.\ \eqref{eq:tn} 
   of the corresponding low-energy Heisenberg model, see the main text.  
   The results remain the same for the number of bath sites $n_b=4$ and $n_b=6$.}
   \label{fig:tn}
   \end{center}
\end{figure}

The main plot in Fig.\ \ref{fig:tn} displays the DMFT results (symbols) for the local spin polarization 
$m=|\langle S_i^z + s_i^z \rangle|$ of the HKH model for the localized spins $S=1$ and $S=2$
vs temperature $T$. We fixed $J_{\rm H}=0.3U/S$ and $U=50t$ which corresponds 
to the bare charge gap $\Delta=80t$ and the exchange interaction $J=4t^2/\Delta=0.05t$. We find no difference in the results
obtained for the number of bath sites $n_b=4$ and $n_b=6$ as in both cases a perfect description 
of the dynamical Weiss field is achieved. 

The mean-field theory results of the corresponding low-energy Heisenberg model obtained from the solution of 
Eq. \eqref{eq:mft} are included in the main plot of Fig.\ \ref{fig:tn} (lines) for comparison.
The main plot in Fig.\ \ref{fig:tn} indicates perfect agreement between the DMFT solution of the HKH model 
and the mean-field theory solution of the corresponding low-energy Heisenberg model for the local spin polarization
corroborating our expectations and arguments on the various kinds of fluctuations.

As the exchange interaction $J$ is increased, the high-energy excitations present in the HKH model \eqref{eq:hkh} 
and neglected in the Heisenberg model \eqref{eq:hem} start to contribute to the thermal fluctuations more and more. 
This explains the observed deviation of the DMFT transition temperature from Eq.\ \eqref{eq:tn} in the inset of 
Fig.\ \ref{fig:tn}.
The precise location of $T_{\rm N}$ is extracted by performing a square root fit 
as expected for mean-field theories to the local magnetization data close to the 
transition temperature. The figure indicates a deviation of $T_{\rm N}$ from Eq.\ \eqref{eq:tn} for $J> 0.16t$. 
This deviation is larger for larger localized spin $S$.

The origin of this deviation is not only the charge excitations but also the spin excitations with the latter playing even
a more important role. In the Heisenberg model \eqref{eq:hem} the size of the local spin is always $\mathcal{S}=S+1/2$. 
The HKH model \eqref{eq:hkh}, however, involves local spin states with the total spin $S+1/2$ and $S-1/2$. The local spin 
excitation is given by $(2S+1)J_{\rm H}=0.3\Delta(2S+1)/1.6S<\Delta$ since we fixed $J_{\rm H}=0.3U/S$. 
The ratio of the local spin excitation to the 
N\'eel temperature \eqref{eq:tn} at $J=0.2t$ equals to $\approx 2.5$ for $S=2$ and $7.5$ for $S=1$. This explains the 
larger deviation from the mean-field Heisenberg results for $S=2$ compared to $S=1$ which we observe in the inset of Fig.\ \ref{fig:tn}. 
In addition, we point out that due to the intersite couplings the actual excitations of the system are dispersive 
and smaller than the local excitations.

The HKH model \eqref{eq:hkh} takes into account the spin excitations of order of
$J_{\rm H}$ present in multi-orbital Mott systems only partially. 
As one can see, for example, for the three-orbital case sketched in the left panel of Fig.\ \ref{fig:hkh}
the Hilbert space of the singly occupied orbitals is given by
\bearr
\label{eq:ss}
\mathcal{H}=\frac{1}{2}\otimes \frac{1}{2}\otimes \frac{1}{2}&=&(1\oplus 0)\otimes \frac{1}{2} \ ,
\eearr
while the HKH model in the right panel of Fig.\ \ref{fig:hkh} involves only the states $1\otimes \frac{1}{2}$, and 
the state $0\otimes \frac{1}{2}$ is neglected.
In the case of more orbitals there are more excited spin states which are neglected by describing a multi-orbital system by the HKH model.
These neglected excited states have an energy gap of order of $(2S+1)J_{\rm H}$ and as long as $(2S+1)J_{\rm H} \gg T$
holds, they are indeed irrelevant and do not contribute substantially to the thermal fluctuations.

\begin{figure}[t]
   \begin{center}
 \includegraphics[width=0.88\columnwidth,angle=-90]{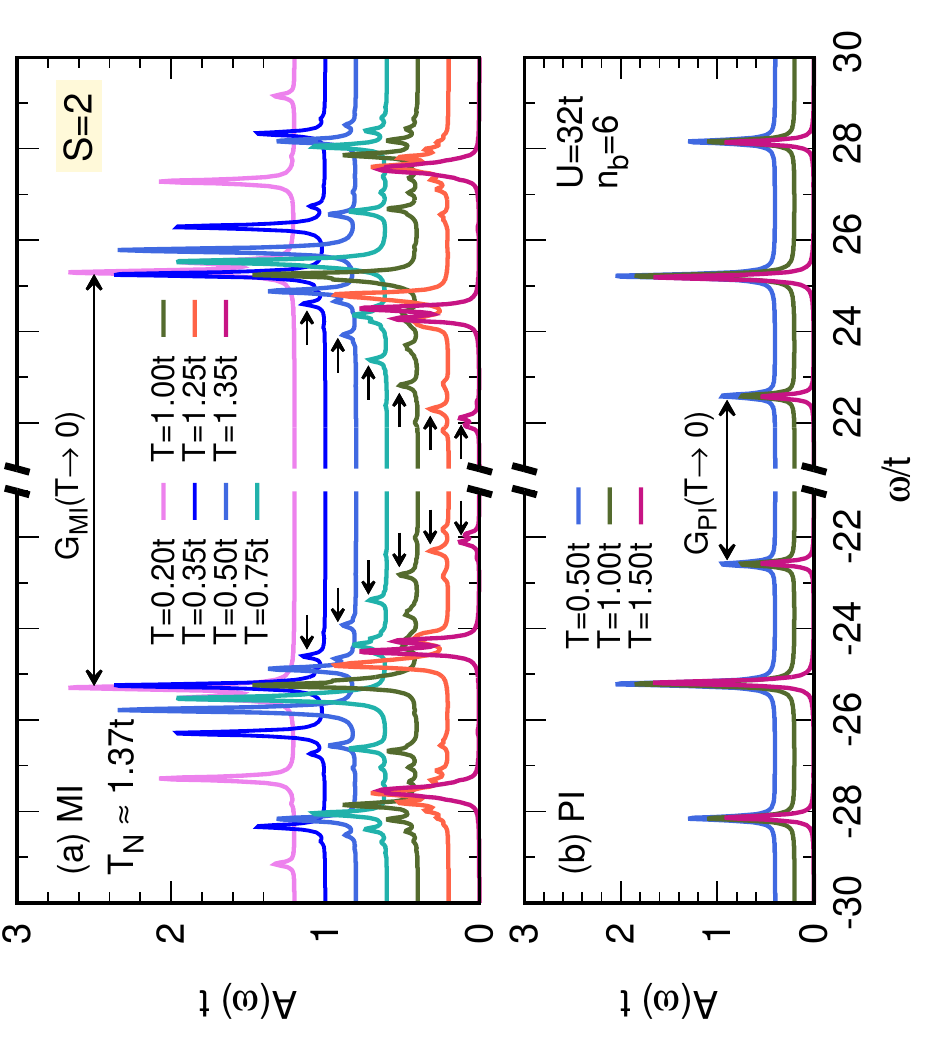}
   \caption{The spectral function $A(\omega)$ vs frequency in the magnetic insulator (MI) phase (a) and 
   in the paramagnetic insulator (PI) phase (b) for the localized spin $S=2$. The Hubbard interaction is given by $U=32t$ 
   and the Hund coupling by $J_{\rm H}=0.3U/S$ implying the bare charge gap $\Delta=1.6U$. 
   The results are for the number of bath sites $n_b=6$. The spectral functions for the different temperatures $T$ are shifted 
   vertically for clarity. The N\'eel temperature is given by $T_{\rm N}\approx 1.37t$. The black arrows indicate the 
   peaks from which the Mott gap in the MI phase $G_{\rm MI}(T)$
   and in the PI phase $G_{\rm PI}(T)$ are read off.}
   \label{fig:sf:s2}
   \end{center}
\end{figure}

The inset of Fig.\ \ref{fig:tn} in fact signals that for $J> 0.16t$ 
the high-energy excitations start to contribute to thermal fluctuations and neither the Heisenberg model \eqref{eq:hem} 
nor the HKH model \eqref{eq:hkh} provide a fully appropriate description of multi-orbital Mott systems at temperatures 
$T \approx T_{\rm N}$ for magnetic exchange couplings of this size. But, for $J< 0.16t$ the thermal fluctuations are solely 
due to the low-energy excitations included in both the Heisenberg model
and the HKH model. In the following, we limit our disscusion to $J< 0.16t$ corresponding to $\Delta>25t$ 
where the HKH model is perfectly justified and examine how 
the low-energy excitations, which are fully taken into account, influence the Mott gap.

Although for the low-energy properties such as the N\'eel temperature a simple mean-field treatment 
of the Heisenberg model \eqref{eq:hem} leads to precisely the same results as the DMFT of the HKH 
model \eqref{eq:hkh} we emphasize that for accessing the charge gap it is essential to 
go beyond the Heisenberg model and the static mean-field approximation and address the HKH model with the DMFT.

\subsection{Spectral redistribution and magnetic blue shift}
\label{sec:mbs}

In this section we examine how the formation of the magnetic ordering below $T_{\rm N}$ affects 
the charge excitations in multi-orbital Mott insulators. We present data for the spectral function 
in Eq.\ \eqref{eq:asf} for the cases $S=1$ and $S=2$ and compare the results with the 
results obtained for the single-orbital Hubbard model in previous studies. This allows us 
to identify the influence of the number of orbitals on the coupling between the charge gap
and the magnetic ordering.

\begin{figure}[t]
   \begin{center}
 \includegraphics[width=0.88\columnwidth,angle=-90]{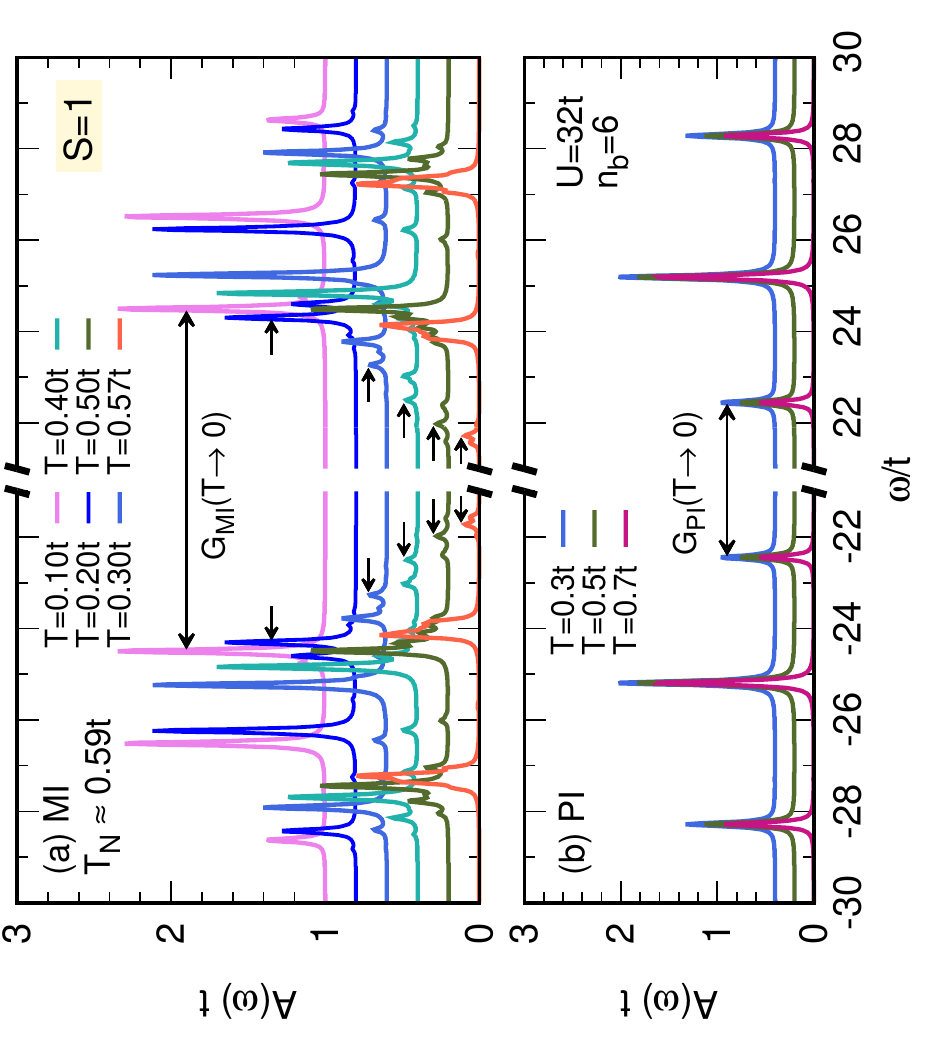}
   \caption{The same as Fig.\ \ref{fig:sf:s2} but for the localized spin $S=1$ with 
   the N\'eel temperature $T_{\rm N}\approx 0.59t$.}
   \label{fig:sf:s1}
   \end{center}
\end{figure}

Fig.\ \ref{fig:sf:s2} displays the spectral function $A(\omega)$ vs frequency $\omega$ for the HKH model \eqref{eq:hkh} 
at different temperatures for the localized spin $S=2$, and Fig.\ \ref{fig:sf:s1} depicts the same for the localized spin $S=1$. 
The Hubbard 
interaction is given by $U=32t$ and the Hund coupling by $J_{\rm H}=0.3U/S$ leading to the
bare charge gap $\Delta=U+2SJ_{\rm H}=1.6U$. 
The results in the MI phase (a) and in the PI phase (b) are shown separately.
Below the N\'eel temperature $T_{\rm N}$, the MI is the stable phase but the metastable PI phase can also be studied by enforcing 
a paramagnetic solution of the DMFT equations. 
We keep $J_{\rm H}S={\rm const.}$ to acquire a 
bare charge gap independent from $S$. This allows us to study the effect which is solely due to the size of
the localized spin on the MBS. 
This procedure implies to use different $J_\text{H}$ for different spin sizes.

It is evident from both figures that the spectral function remains essentially unchanged in the PI phase upon reducing 
the temperature from the N\'eel temperature $T_{\rm N}$ down to zero. 
Including non-local spin fluctuations beyond DMFT may approach the spectral functions
in the PI phase closer to the one in the MI phase.  
However, these contributions are expected to be negligible in systems with large 
coordination numbers where the DMFT is justified. In contrast to the PI phase,
there is a shift of
the electron ($\omega>0$) and the hole ($\omega<0$) peaks denoted by arrows away from the Fermi 
energy $\omega=0$ in the MI phase. This leads to an MBS of the Mott gap. We find no change in the spectral function in the MI phase
below $T=0.2t$ for $S=2$ and below $T=0.1t$ for $S=1$ meaning that these temperatures can be viewed as equivalent to zero temperature. 
Although the results are obtained for the finite number of bath sites $n_b=6$ in the ED impurity solver, we observe a very similar 
behavior also for $n_b=4$ shown in Appendix \ref{app:A}. The shift of the electron and hole spectral peaks 
towards higher energies due to antiferromagnetic ordering is already realized in the single-orbital Hubbard model \cite{Sangiovanni2006,Wang2009,Fratino2017}.
Our findings in Figs. \ref{fig:sf:s2} and \ref{fig:sf:s1} extend these studies to the multi-orbital case. 

Due to the finite number of bath sites the spectral function is not smooth but consists of separate sharp peaks
whose weight distribution approximate the one of the continuous spectral function.
Nevertheless, one can still see from Figs. \ref{fig:sf:s2} and \ref{fig:sf:s1} how the spectral function 
in the MI phase changes upon increasing the temperature 
from $0$ to $T_{\rm N}$ and takes a similar form as the spectral function in the PI phase. 
For continuity, one expects the spectral function in the MI and in the PI to be exactly the same as $T \to T_\text{N}$.
The main difference is that at $T\approx T_{\rm N}$ the electron and the hole spectral contributions in 
the MI phase are shifted further towards the Fermi energy in contrast to the PI phase.
This slight inconsistency originates from the finite number of bath sites 
used in the ED solver as well as from the generally accepted fact that achieving accurate
results near a transition point is demanding even for static quantities, let alone dynamical functions. 
This suggests that the spectral function in the PI phase should be considered more accurate for temperatures 
close to $T_{\rm N}$.  

\begin{figure}[t]
   \begin{center}
   \includegraphics[width=0.95\columnwidth,angle=0]{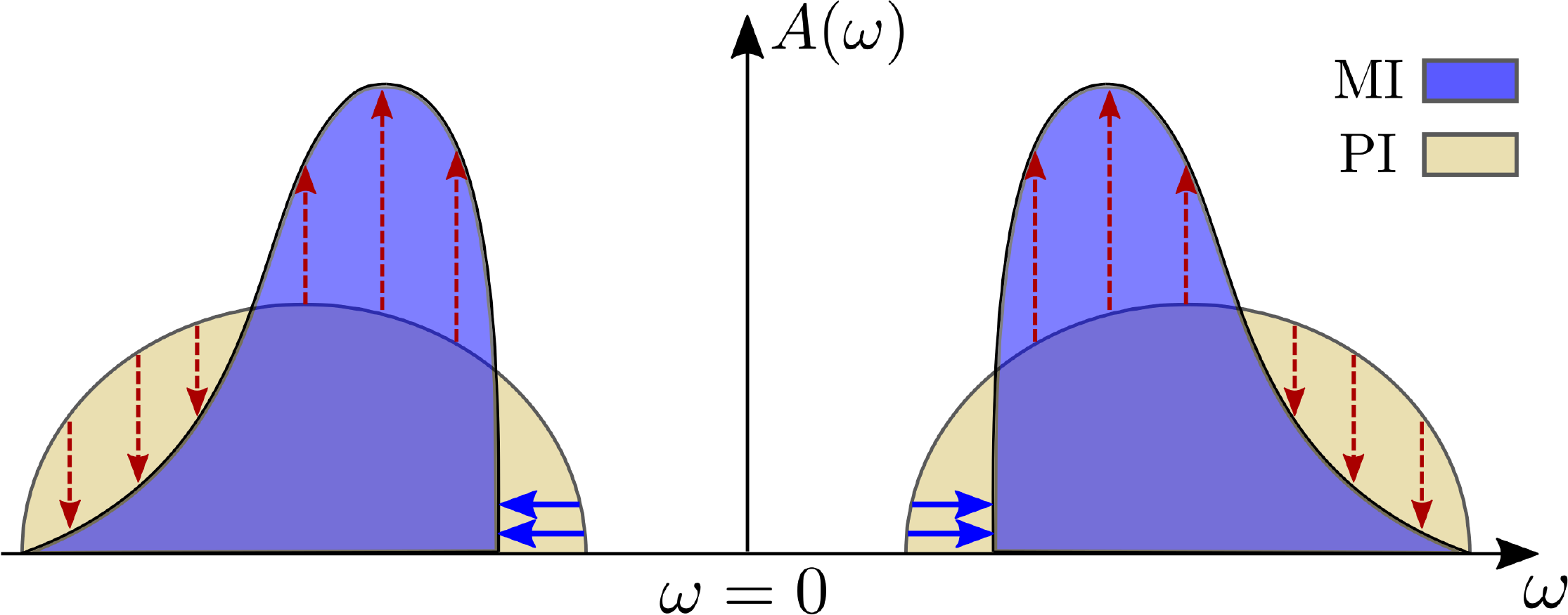}
   \caption{Schematic representation of the spectral function $A(\omega)$ in the 
   paramagnetic insulator (PI) and in the magnetic insulator (MI) at the temperature
   $T\to 0$. The magnetic blue shift of the Mott gap is denoted by continuous blue arrows 
   and the magnetic redistribution of spectral weight by dashed red arrows.}
   \label{fig:mse}
   \end{center}
\end{figure}

One can see from Figs. \ref{fig:sf:s2} and \ref{fig:sf:s1} that for $T\to 0$ the low-energy electron and hole peaks, 
indicated by arrows, carry a spectral weight which is larger in the MI phase than in the PI phase. 
Such a difference between the spectral weight distribution in the MI and in the PI
phases was first pointed out for the single-orbital Hubbard model in Ref.\ \cite{Sangiovanni2006}. 
It was found qualitatively consistent with photoemission spectroscopy measurements on Cr-doped V$_2$O$_3$ \cite{Sangiovanni2006}. 
In Ref.\ \cite{Hafez-Torbati2021} we demonstrated that such a magnetic redistribution of the spectral weight 
would result in an {\it increase} in the free energy unless the effect is compensated by an MBS  
of the Mott gap. One has to keep in mind that a decrease in the free energy is the prerequisite 
for the MI phase to stabilize. The MBS of the Mott gap (blue arrows) and the magnetic redistribution of 
spectral weight (red arrows) are summarized schematically in Fig.\ \ref{fig:mse}.
The fact that the magnetic redistribution of the spectral weight and the MBS of the Mott gap occur in 
multi-orbital systems, see Figs. \ref{fig:sf:s2} and \ref{fig:sf:s1}, very similar to the single-orbital 
case \cite{Hafez-Torbati2021} corrobarates the view that they are generic spectral consequences of antiferromagnetic ordering 
in Mott insulators. This view is further supported by the connection to the decrease of the free energy \cite{Hafez-Torbati2021}.

So far, we discussed the effects common to the representative values $S=0$, $S=1$, and $S=2$. 
Next, we turn to the differences between the results for these cases which indicate the 
general trends for varying $S$, i.e., the number of orbitals $2S+1$.
The study of these differences is essential in order to identify Mott insulators 
representing a maximum coupling between the magnetic order and the electronic structure. 

As one can see from Figs.\ \ref{fig:sf:s2}(b) and \ref{fig:sf:s1}(b) the electron and the hole 
peaks defining the gap $G_{\rm PI}(T)$ show no strong dependence on the size of the localized 
spin $S$ although for the $S=2$ case the gap seems a bit larger than for the $S=1$ case. However, in the MI phase 
the gap $G_{\rm MI}(T)$ as $T\to 0$ is clearly larger for the $S=2$ case in 
Fig. \ref{fig:sf:s2}(a) in contrast to the $S=1$ case in Fig.\ \ref{fig:sf:s1}(a). 

\begin{figure}[t]
   \begin{center}
 \includegraphics[width=1.1\columnwidth,angle=-90]{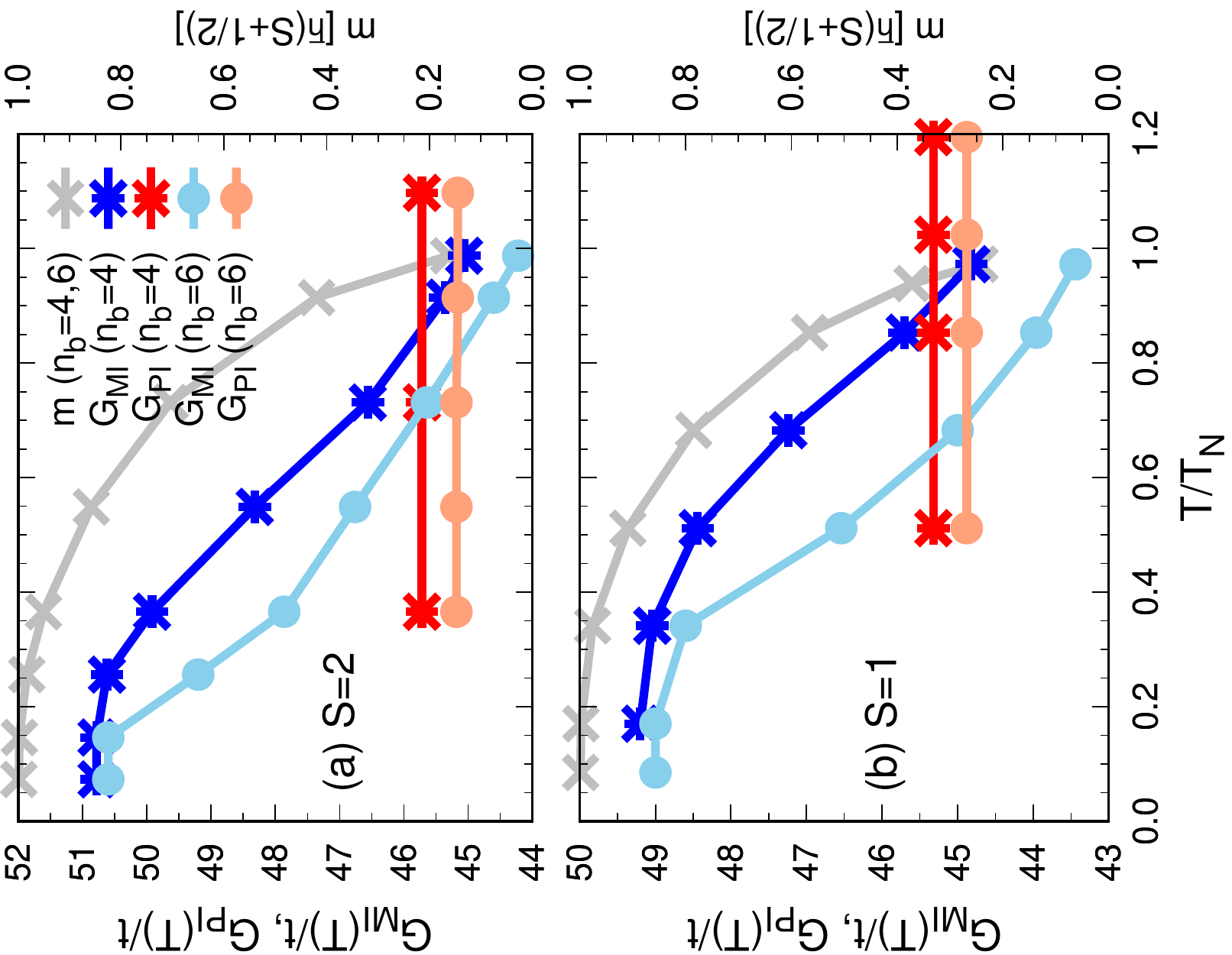}
   \caption{The Mott gap in the magnetic insulator (MI) phase $G_{\rm MI}(T)$ and 
   in the paramagnetic insulator (PI) phase $G_{\rm PI}(T)$ plotted vs temperature $T$ for the localized spin $S=2$ (a) and $S=1$ (b).
   We set the Hubbard interaction to $U=32t$ and the Hund coupling to $J_{\rm H}=0.3U/S$. The 
   bare charge gap is given by $\Delta=U+2SJ_{\rm H}=1.6U$. The figure also includes the 
   local spin polarization $m=|\langle S_i^z+s_i^z \rangle|$, see right axis for its scale.
   The results for the number of bath sites $n_b=6$ and $n_b=4$ are compared.}
   \label{fig:gap}
   \end{center}
\end{figure}

To investigate the gap in more details and to reveal the dependence of the MBS in Eq.\ \eqref{eq:mbs} on $S$, 
we plot the gap in the PI phase $G_{\rm PI}(T)$ and in the MI phase $G_{\rm MI}(T)$ 
as function of $T$ in Fig.~\ref{fig:gap} for $S=2$ (a) and for $S=1$ (b). 
The local spin polarization $m$, right axis, is also shown. 
The figures contain the results for the number of bath sites $n_b=6$ and $n_b=4$. 
The gap is extracted from the positions of the peaks closest to the Fermi energy specified by arrows 
for $n_b=6$ in Figs. \ref{fig:sf:s2} and \ref{fig:sf:s1} and for $n_b=4$ in Figs. \ref{app:fig:sf:s2}
and \ref{app:fig:sf:s1} in Appendix \ref{app:A}.

Fig. \ref{fig:gap} shows that the gap in the PI phase remains 
essentially unchanged as function of temperature while in the MI phase the gap increases and the spin
polarization grows upon $T\to0$ for both $n_b = 6$ and $n_b = 4$.
The fact that the gaps in the MI phase and in the 
PI phase do not continuously join at the N\'eel temperature originates from the finite number of bath sites in the ED solver
as we discussed above for the spectral function.

For temperatures near zero in Fig. \ref{fig:gap} we find a nice agreement 
between the results for $n_b=4$ and $n_b=6$ which permits an accurate estimate of the gap and an investigation
of the effect of the size of the localized spin on the MBS. The PI gap $G_{\rm PI}(T)$
for the localized spin $S=1$ in Fig.\ \ref{fig:gap}(b) is around $45t$ while for the localized spin $S=2$ 
in Fig.\ \ref{fig:gap}(a) it is around $45.5t$. The MI gap $G_{\rm MI}(T)$ at $T\to 0$ is around $49.1t$ for $S=1$ 
in Fig.\ \ref{fig:gap}(b) and it is around $50.7t$ for $S=2$ in Fig.\ \ref{fig:gap}(a). These gap values correspond 
to an MBS $\Gamma_{\rm MG}(0)\approx 4.1t$ for $S=1$ and an MBS $\Gamma_{\rm MG}(0)\approx 5.2t$ for $S=2$ which are significantly 
larger than the MBS $\Gamma_{\rm MG}(0)\approx 1.1t$ we found in Ref.\ \cite{Hafez-Torbati2021} for the single-orbital
case ($S=0$) with the same bare charge gap $\Delta \approx 50t$. 
Hence, we arrive at the summarizing conclusion that increasing the size of the localized spin
increases the Mott gap in both the PI phase and the MI phase.
But, the increase in the MI phase is larger leading eventually to a larger MBS for larger localized spin or, equivalently,
larger number of involved orbitals.

\subsection{Hopping and exchange contributions to\\magnetic blue shift}
\label{sec:he}

The double exchange mechanism is well-known since 1950s to be responsible for the ferromagnetic ordering
in doped perovskite manganites. 
The strong ferrromagnetic Hund coupling (Hund's first rule) forces the spin of the itinerant electron to be aligned with the 
localized spin at each lattice site. Hence, the electron motion is strongly hindered if the adjacent local spins are 
antiparallel while parallel orientation allows the electron to move
\cite{Zener1951,Anderson1955,Gennes1960,Pavarini2012,Hartmann1989}.

In Ref.\ \cite{Hafez-Torbati2021} we revealed that in ferromagnetic Kondo lattice systems the double exchange mechanism 
reduces the effective hopping upon transition from a paramagnetic insulator to an antiferromagnetic 
insulator, which narrows the spectral bandwidth and results 
in an MBS which is enhanced relative to its value in a pure Hubbard model.
For the Hubbard-Kondo model consisting of the terms \eqref{eq:hu} and \eqref{eq:ko}, 
\be
\label{eq:hk}
H_{\rm HK}=H_{\rm Hu}+H_{\rm Ko}, 
\ee
with a localized spin $S=2$ and a bare charge gap $\Delta \approx 50t$, we identified an MBS $\Gamma_{\rm MG}(0)\approx 3.2t$
larger than $\Gamma_{\rm MG}(0)\approx 1.1t$ in the single-orbital Hubbard model \eqref{eq:hu} alone. Nevertheless, the MBS 
$\Gamma_{\rm MG}(0)\approx 3.2t$ is still considerably smaller than $\Gamma_{\rm MG}(0)\approx 5.2t$ as we found for the full
HKH model 
\eqref{eq:hkh} with the same localized spin $S=2$. 
Since the double exchange mechanism plays the same role in the HK model as in the HKH model we attribute this additional 
enhancement to the exchange interaction \eqref{eq:he}.

\begin{figure}[t]
   \begin{center}
 \includegraphics[width=1.15\columnwidth,angle=-90]{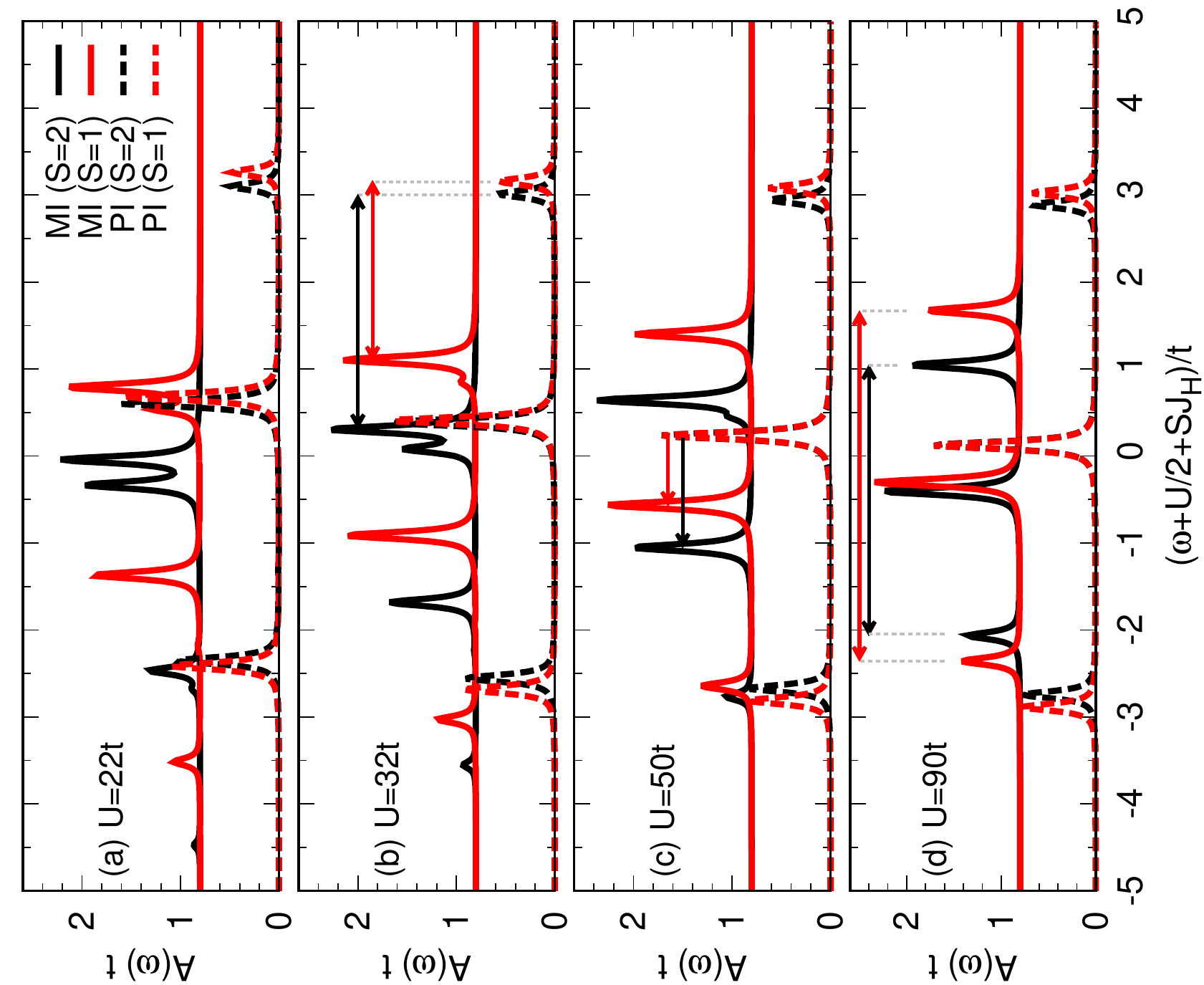}
   \caption{The spectral function $A(\omega)$ vs frequency $\omega$ at $T=0$ for various values of the Hubbard $U$. 
   The bare charge gap is given by $\Delta=U+2SJ_{\rm H}=1.6U$ which corresponds to the Hund coupling $J_{\rm H}=0.3U$ for the localized spin $S=1$ and $J_{\rm H}=0.15U$ for $S=2$.
   The results are for the number of bath sites $n_b=6$.
   The frequency range is restricted to the range of the hole contribution of the spectral function. 
   The electron contribution is the mirror image of the hole contribution with respect to
   the Fermi energy $\omega=0$. At each panel the results for the magnetic insulator (MI) phase and for the 
   paramagnetic insulator (PI) phase with different localized spins $S=2$ and $S=1$ are shown for comparison.
   The results in the MI phase are shifted vertically for clarity.
   The arrows are used in panel (b) to illustrate the magnetic blue shift, in panel (c) to show the shift 
   of the spectral function in the MI phase relative to the PI phase, and in panel (d) to specify the spectral 
   bandwidth in the MI phase, see the main text.}
   \label{fig:sf:t0}
   \end{center}
\end{figure}

To unveil the role of the exchange interaction and of the double exchange mechanism in the MBS in the HKH model 
we plot the hole contribution ($\omega<0$) of the spectral function in Fig.\ \ref{fig:sf:t0} at $T=0$ for 
different values of $U$. The bare charge gap is given by $\Delta=U+2SJ_{\rm H}=1.6U$. 
The spectral functions in the PI and MI phase for localized spins $S=1$ and $S=2$ are shown for comparison. The results in the MI phase are shifted vertically 
for clarity. We display only the hole contribution for a better visibility of the differences. The 
electron contribution is the mirror image of the hole contribution with respect to the Fermi energy $\omega=0$
due to the electron-hole symmetry of the studied model. This symmetry of the HKH model is always perfectly realized in our results as one can see, for instance, 
from Figs. \ref{fig:sf:s2} and \ref{fig:sf:s1}. 

The difference between the peaks closest to the Fermi energy (the rightmost peaks) in the MI phase and in the PI 
phase in Fig.\ \ref{fig:sf:t0} equals half of the MBS, $\Gamma_{\rm MG}(0)/2$. This difference is illustrated by double-head 
arrows in panel (b) of Fig.\ \ref{fig:sf:t0} for the different localized spins $S$. 
The message of Fig.\ \ref{fig:sf:t0} is very clear: Increasing the bare charge gap $\Delta$ 
from Fig.\ \ref{fig:sf:t0}(a) to Fig.\ \ref{fig:sf:t0}(d) decreases the MBS. Concerning the dependence on $S$ the larger $S$
entails a larger MBS.

One can identify two main origins behind the MBS, which become more and more distinct as $\Delta$ becomes larger
in Fig.\ \ref{fig:sf:t0}. 
One contribution is a shift of the spectral function in the MI phase 
towards higher energies (away from the Fermi energy) relative to the spectral function in the PI phase. 
This can be seen especially from Fig.\ \ref{fig:sf:t0}(c) and Fig.\ \ref{fig:sf:t0}(d) where
the middle peak of the spectral function in the MI phase is shifted with respect to the middle peak of 
the spectral function in the PI phase. This shift is indicated in Fig.\ \ref{fig:sf:t0}(c) by arrows and is larger for the 
larger localized spin and it decreases upon increasing the bare charge gap $\Delta$. A similar, but smaller shift 
of the spectral function can be seen in our results for the single-orbital Hubbard model and for the Hubbard-Kondo model
in Ref.\ \cite{Hafez-Torbati2021}. This shift stems from the effective antiferromagnetic exchange interaction
which results from the second order lowering in energy by the virtual hopping of a spin-$\uparrow$ electron onto a site 
with a spin-$\downarrow$ electron or vice versa. For parallel spins this lowering does not occur due to Pauli's principle \cite{Pavarini2012,Fazekas1999}. 
The exchange interaction can be described in the MI phase by a mean-field effective magnetic field with the strength
\be
\label{eq:hi:tot}
h_i^{\vpdag}=6J\langle S^z_i+s^z_i \rangle
\ee
at the lattice site $i$. One notes that the term $6J\langle s^z_i \rangle$, absent in \eqref{eq:hi:iti}, 
is due to the itinerant Hamiltonian \eqref{eq:hu}. The effective magnetic field \eqref{eq:hi:tot} induces 
a shift of the electron and the hole spectral contributions away from the Fermi energy in the MI phase. 
The relation \eqref{eq:hi:tot} explains the larger shift of the spectral function we observe for the larger 
localized spin, and the smaller shift we observe for larger $\Delta$ because it corresponds to a smaller $J=4t^2/\Delta$.
We refer to this contribution to the MBS as the exchange contribution.

\begin{figure}[t]
   \begin{center}
 \includegraphics[width=0.85\columnwidth,angle=-90]{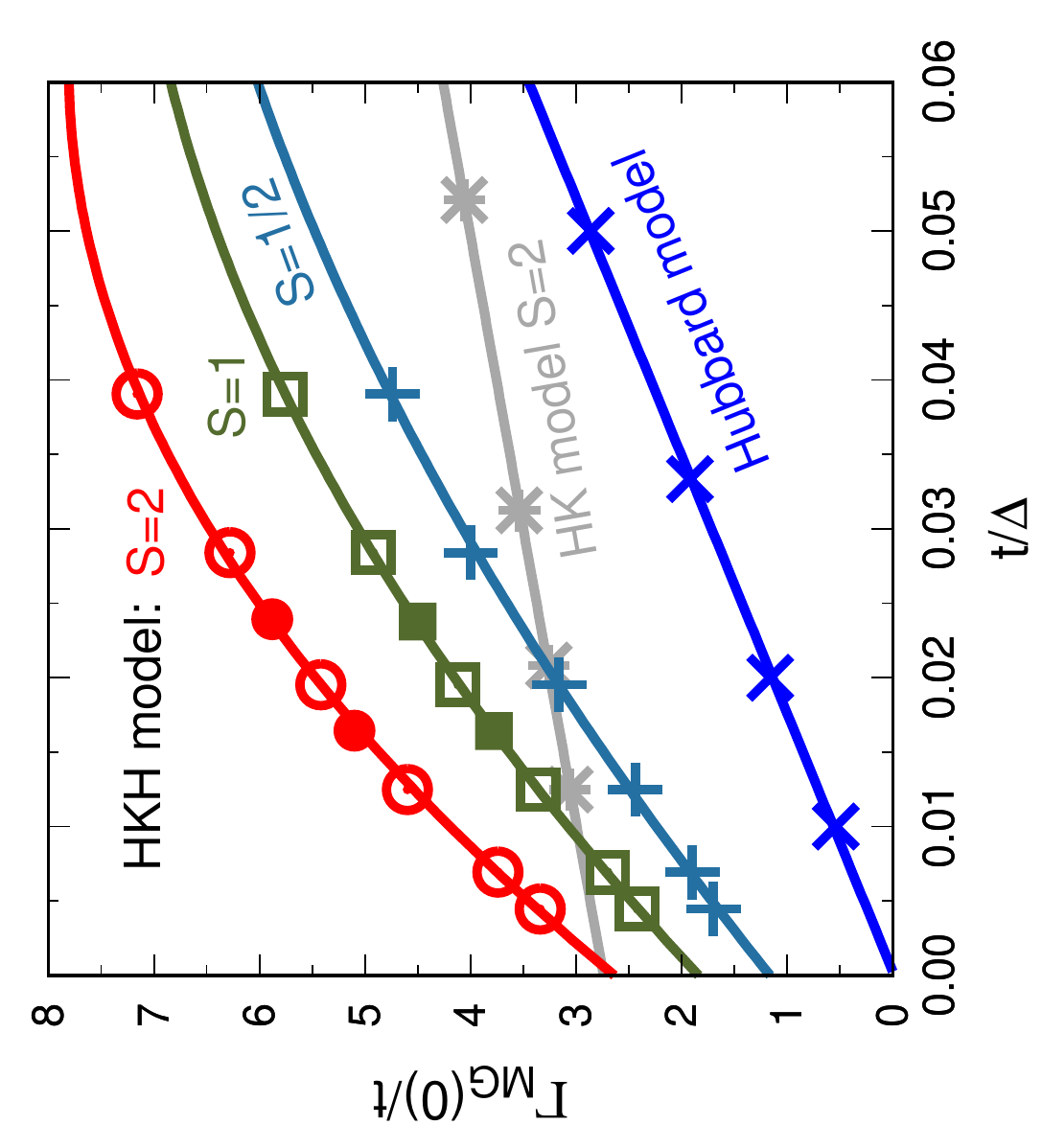}
   \caption{The magnetic blue shift $\Gamma_{\rm MG}(T)$ given by Eq.\ \eqref{eq:mbs} at $T=0$ vs the inverse bare charge gap $\Delta=U+2SJ_{\rm H}$
   in the Hubbard-Kondo-Heisenberg (HKH) model \eqref{eq:hkh} with the localized spin $S=\frac{1}{2},1,2$. The 
   hopping parameter $t$ is used as the unit of energy on both axes. The results for the Hubbard-Kondo (HK) model with 
   $S=2$ and for the Hubbard model from Ref.\ \cite{Hafez-Torbati2021} are included for comparison. The results 
   are obtained for $J_{\rm H}=0.3U/S$ varying $U$, except for the filled circles (HKH model with $S=2$) and the filled squares 
   (HKH model with $S=1$) which are for $J_{\rm H}=0.225U$ independent from $S$.}
   \label{fig:mbs}
   \end{center}
\end{figure}

The other contribution to the MBS originates from a narrower spectral bandwidth in the MI phase in contrast to the 
spectral bandwidth in the PI phase due to the double exchange mechanism. 
This can best be seen in panel (d) in Fig.\ \ref{fig:sf:t0}. 
In this panel we exemplarily depict the spectral bandwidth in the MI phase for $S=1$ and $S=2$ by double-head arrows.
In Ref.\ \cite{Hafez-Torbati2021} we discussed how the double exchange mechanism shrinks the spectral bandwidth 
upon transition from the PI to the MI phase leading to a contribution proportional to the hopping to the MBS. We 
refer to this contribution as the double exchange contribution or the hopping contribution. 

The additional aspect one can read off from Fig.\ \ref{fig:sf:t0} is that how the hopping contribution 
depends on the size of the localized spin $S$. Upon increasing the size of the localized spin from $S=1$ to $S=2$ the spectral bandwidth 
in both the PI phase and the MI phase decreases. However, this decrease is larger in the MI phase than in the PI phase
which results in a larger hopping contribution to the MBS for larger localized spin.

In order to extract quantitative values of the hopping and the exchange contributions to the MBS for the 
localized spins $S=\frac{1}{2},1,2$ in the HKH model \eqref{eq:hkh} we plot 
the MBS $\Gamma_{\rm MG}(0)$ vs the inverse bare charge gap $\Delta$ in Fig.\ \ref{fig:mbs}. 
The hopping parameter $t$ is used as the unit of energy on both axes. 
A quadratic fit shown by a continuous line nicely describes the data leading to the following expansion for 
the MBS in the HKH model in powers of $J/t=4t/\Delta$,
\bearr
\label{eq:mbs:exp}
\frac{\Gamma_{\rm MG}(0)}{t}&=&C_0 + C_1 \frac{4t}{\Delta} + C_2 \frac{16t^2}{\Delta^2} + \cdots \nn \\
&=&C_0 + C_1 \frac{J}{t} + C_2 \frac{J^2}{t^2} + \cdots \ .
\eearr
One can see from Eq.\ \eqref{eq:mbs:exp} that there is a hopping contribution $\Gamma_{\rm MG}^{(t)}(0)=C_0t$ and an exchange contribution 
$\Gamma_{\rm MG}^{(J)}(0)=C_1J$ to the MBS. The hopping contribution is proportional to the offset and the exchange contribution is proportional
to the slope in Fig.\ \ref{fig:mbs} in the limit $t/\Delta \to 0$. 
The expansion coefficients in Eq.\ \eqref{eq:mbs:exp} are provided in Table \ref{tab:ec} for different spin $S$.
Note that Table \ref{tab:ec} provides the expansion coefficients also for the $S=3/2$ case. The data for $S=3/2$ are not 
included in Fig.\ \ref{fig:mbs} in order not to overload the figure.
In Fig.\ \ref{fig:mbs}, we also included our results for the Hubbard model \eqref{eq:hu} 
and the HK model \eqref{eq:hk} with the localized spin $S=2$ from Ref.\ \cite{Hafez-Torbati2021} and their linear fit 
for comparison. 
The results in Fig.\ \ref{fig:mbs} are obtained for $J_{\rm H}=0.3U/S$ varying $U$, except for the filled circles (HKH model with $S=2$) and
the filled squares (HKH model with $S=1$) which are computed for a fixed $J_{\rm H}=0.225U$ independent from $S$. 
The fit is performed for the data with $J_{\rm H}=0.3U/S$ but nicely describes also the data 
obtained for $J_{\rm H}=0.225U$. This confirms that the MBS depends only on the bare charge gap $\Delta=U+2SJ_{\rm H}$ and 
not on $J_{\rm H}$ and $U$ individually.

\begin{table}[t]
\begin{center}
    \begin{tabular}{ c | c | c | c | c | c }
          & $S=0$ & $S=1/2$  & $S=1$   & $S=3/2$ & $S=2$ \\ \hline
    $C_0$ & $0$    & $1.16$  & $1.84$ & $2.30$ & $2.64$ \\ \hline
    $C_1$ & $14.4$ & $28.3$ & $33.0$ & $38.5$ & $42.4$ \\ \hline
    $C_2$ & $0$    & $-33.5$ & $-50.4$ & $-73.6$ & $-86.8$ \\
    \end{tabular}
\end{center}
 \caption{The expansion coefficients for the MBS of the Mott gap in Eq.\ \eqref{eq:mbs:exp}. The results are for
 the HKH model in Eq.\ \eqref{eq:hkh} with the localized spin $S$. One notes that the localized spin $S$ in the HKH model 
 corresponds to a Mott insulator with the number of orbitals $2S+1$.}
\label{tab:ec}
\end{table}

One can see from Table \ref{tab:ec} that the hopping $C_0$ and the exchange $C_1$ coefficients both increase 
upon increasing the size of the localized spin in the HKH model \eqref{eq:hkh}. 
One notes that there is also a negative quadratic contribution $C_2$ which becomes larger 
for larger localized spins and suppresses the MBS as $t/\Delta$ increases. 
The exchange coefficient $C_1=28.3$ for $S=1/2$ is almost just twice 
the exchange coefficient $C_1=14.4$ for $S=0$. This is nicely consistent with our explanation of the origin of 
the exchange contribution of the MBS based on the mean-field effective magnetic field in Eq.\ \eqref{eq:hi:tot}
which suggests a proportionality of $C_1$ to $S+1/2$. 
For larger localized spins, however, the exchange coefficient starts to deviate from the value $(2S+1)\times 14.4$.
We associate such a deviation to higher order terms beyond the mean-field effective magnetic field \eqref{eq:hi:tot} 
which become more and more important for larger localized spins.

An equal hopping contribution in the HKH model \eqref{eq:hkh} and in the HK model \eqref{eq:hk} having the same 
localized spin $S$ is expected since in the limit $t/\Delta \to 0$ the Heisenberg term \eqref{eq:he} vanishes 
(since we have fixed $J=4t^2/\Delta$ in Eq.\ \eqref{eq:he}). 
Indeed, Fig.\ \ref{fig:mbs} shows a nice agreement between the hopping contribution in the HKH model ($C_0\approx 2.64$) and in 
the HK model ($C_0\approx 2.74$) for the localized spin $S=2$. 
This provides a corroborating consistency test for our results. 
As $t/\Delta$ is increased from zero in Fig.\ \ref{fig:mbs}, the MBS in the HKH model with $S=2$ increases much faster 
than the MBS in the HK model due to the additional exchange interaction \eqref{eq:he}.

\section{Application}
\label{sec:app}

The expansion for the MBS of the Mott gap in Eq.\ \eqref{eq:mbs:exp} is obtained for the simple cubic lattice 
with the hopping and the exchange interactions limited to nearest-neighbor sites. In real materials, 
however, one needs to deal with various kinds of lattice structures and long-range terms beyond nearest
neighbor are commonly involved. 
The aim of this section is to generalize Eq.\ \eqref{eq:mbs:exp} such that it can be employed to 
estimate the MBS for a larger variety of antiferromagnetic compounds.

As we discussed in the previous section, one contribution to the MBS originates from the exchange interaction
and can be understood based on the effective mean-field magnetic field in Eq.\ \eqref{eq:hi:tot}. 
This suggests to substitute the exchange $J$ in Eq.\ \eqref{eq:mbs:exp}, which holds for $Z=6$ neighbors,
by an effective value~$\tilde J$
\be
\label{eq:gj}
ZJ \longrightarrow Z\tilde{J}=\sum_i \sigma_i Z_i J_i
\ee
for a system with exchange interactions represented by $J_i$ for the $i$th neighbor and the corresponding 
coordination number $Z_i$.  The sign factor $\sigma_i= \pm 1$ depends on the spin orientation on the lattice sites linked 
by $J_i$. It takes the positive value $+1$ if the sites linked by $J_i$ 
have antiparallel spin ordering and the negative value $-1$ if the sites linked by $J_i$ have parallel spin 
ordering. 
This is because an antiferromagnetic exchange interaction $J_i>0$ linking sites with antiparallel 
spin ordering or a ferromagnetic exchange interaction $J_i<0$ linking sites with parallel spin ordering 
enhance the effective mean-field magnetic field while an antiferromagnetic exchange interaction 
linking sites with parallel spin ordering or a ferromagnetic exchange interaction linking sites with antiparallel 
spin ordering suppress the effective mean-field magnetic field. 
This accumulates the energetically favored spin alignment in the mean-field theory.

The expansion in Eq.\ \eqref{eq:mbs:exp} is obtained for an ideal antiferromagnetic system, i.e., 
the exchange interaction is always antiferromagnetic linking only sites with antiparallel spin ordering. 
Accordingly, we expect the generalization to be valid for a system which contains no or weak ferromagnetic 
exchange interactions in contrast to the antiferromagnetic ones. In addition, the antiferromagnetic exchange 
interactions linking sites with parallel spin ordering should be weak.

As we discussed in the previous section and also in Ref.\ \cite{Hafez-Torbati2021}, the hopping contribution 
to the MBS stems from the narrowing of the spectral bandwidth by the double exchange mechanism. 
The double exchange mechanism reduces the effective hopping between sites 
with antiparallel spin ordering upon transition from the PI to the MI phase. If the sites have 
parallel spin ordering the double exchange mechanism enhances the effective hopping \cite{Hafez-Torbati2021} 
which is expected to induce a decrease in the MBS. This suggests the substitution of the hopping $t$ in Eq.\ 
\eqref{eq:mbs:exp} by an effective hopping $\tilde t$ according to
\be
\label{eq:gt}
Zt \longrightarrow Z\tilde{t}=\sum_i \sigma_i Z_i t_i
\ee
where $t_i$ is the hopping to the $i$th neighbor given by $t_i=\sqrt{\Delta J_i}/2$ for the antiferromagnetic
exchange interaction $J_i>0$. The bare charge gap is computed from the Hubbard $U$ and the Hund coupling $J_{\rm H}$
as $\Delta=U+2SJ_{\rm H}$. We set $t_i=0$ for the ferromagnetic exchange interaction $J_i<0$ which is 
supposed to be weak. The sign factor $\sigma_i= \pm 1$ has the same definition as above, i.e., it is
$+1$ if $t_i$ is between sites with antiparallel spins and $-1$ if $t_i$ is between sites with parallel spins.

The exchange interactions can be extracted adequately by fitting the inelastic neutron scattering
data of low-energy excitations to the magnon dispersion obtained by means of spin wave theory for instance.
The approach has already be applied to a variety of magnetic materials and the exchange interactions 
are known. 
The atomic physics or the density functional theory calculations yield an estimate of the Hubbard interaction 
$U$ and the Hund coupling $J_{\rm H}$ which determine the bare charge gap $\Delta=U+2SJ_{\rm H}$. Hence, the relation 
\eqref{eq:mbs:exp} with the generalizations \eqref{eq:gj} and \eqref{eq:gt} reads
\be
\label{eq:gmbs}
\Gamma_{\rm MG}(0)=C_0\tilde{t}+C_1\tilde{J}+C_2\frac{\tilde{J}^2}{\tilde{t}} 
\ee
and provides a convenient way of evaluating the MBS of the Mott gap in antiferromagnetic materials.

In the following we apply the method to the room temperature antiferromagnets $\alpha$-MnTe, NiO, and BiFeO$_3$
which are considered as promising candidates \cite{Baltz2018} for application in antiferromagnetic spintronics. These systems 
are better described as charge-transfer insulators \cite{Zaanen1985} rather than Mott insulators. The charge gap, known 
as the charge-transfer gap, is defined between the occupied $p$ band and the empty upper Hubbard band. The MBS of the 
charge-transfer gap can be approximated as half the MBS of the Mott gap \cite{Hafez-Torbati2021},
\be
\Gamma_{\rm CTG}(T) \approx \frac{1}{2}\Gamma_{\rm MG}(T) \ .
\ee
This relies on the assumption that the effect of the magnetic ordering on the $p$ band is negligible as it is only 
indirectly affected by the magnetic ordering, and that the shifts of the upper and the lower Hubbard bands to 
higher energies are equal which is typical for half-filled Mott insulators. For a more detail discussion 
we refer the reader to Ref.\ \cite{Hafez-Torbati2021}.

We point out that we neglect vertex corrections in our calculations of
the one-particle propagator and its spectral density. They enhance the formation
of excitons which influence the charge gap as measured in experiment. But
in DMFT they do not enter in conductivity at all \cite{Khurana1990} and are 
expected to be small anyway for local interactions in large dimensions. 
In lower dimensions, this can be different \cite{Essler2001,Kauch2020}.

\subsection{$\alpha$-MnTe}
\begin{figure}
 \includegraphics[width=0.75\columnwidth]{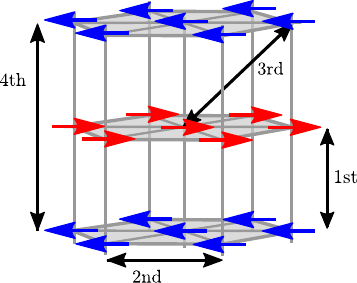}
 \caption{Antiferromagnetic order in $\alpha$-MnTe. There are ferromagnetic triangular layers which are stacked 
 antiferromagnetically. The 1st, 2nd, 3rd, and 4th neighbors are specified.}
 \label{fig:mnte}
\end{figure}

We start with the MBS in $\alpha$-MnTe for which we performed an extensive analysis in Ref.\ \cite{Hafez-Torbati2021}
and for which there is also experimental data available \cite{Ferrer-Roca2000,Bossini2020}. This permits to examine
how well the generalization in Eqs. \eqref{eq:gj} and \eqref{eq:gt} is grounded before proceeding to the other compounds. 
The antiferromagnetic order in $\alpha$-MnTe develops below the N\'eel temperature $T_{\rm N} \approx 310$ K with
the magnetic Mn$^{+2}$ ions forming triangular ferromagnetic layers which are stacked antiferromagnetically. 
The magnetic order and the neighboring sites up to the fourth neighbor are specified in Fig.\ \ref{fig:mnte}. 
The exchange interactions involving up to third neighbor are computed by fitting the magnon dispersion to the 
inelastic neutron scattering data \cite{Szuszkiewicz2006}. In Ref.\ \cite{Mu2019} the exchange interactions are 
slightly modified adding also a fourth neighbor term in order to attain accurate results for the N\'eel temperature.
We use the exchange couplings of Ref.\ \cite{Mu2019} given by $J_1=3.072$ meV, $J_2=0.0272$ meV, $J_3=0.4$ meV, and $J_4=0.16$ meV 
(according to our definition of the exchange interaction in Eq.\ \eqref{eq:hem}) which are all antiferromagnetic 
allowing to assign a hopping term to them. 
One notes that the dominent terms $J_1$ and $J_3$ are between sites 
with antiparallel spin order. The terms $J_2$ and $J_4$ linking parallel spins are weak. 
We consider a Hubbard interaction $U=5.5$ eV and a Hund coupling $J_{\rm H}=0.8$ eV
based on the atomic physics \cite{Bossini2020} and the density functional theory \cite{Mu2019}. The half filled 
$3d$ shell of Mn$^{+2}$ ions corresponds to the localized spin $S=2$ in the HKH model \eqref{eq:hkh}. 
These values result in the hopping parameters $t_1 \approx 81.74$ meV, $t_2\approx7.69$ meV, $t_3\approx29.49$ meV, 
and $t_4\approx18.65$ meV.

Substituting the hopping and the exchange parameters in Eqs. \eqref{eq:gt} and \eqref{eq:gj} with the 
coordination numbers $Z_1=2$, $Z_2=6$, $Z_3=12$, and $Z_4=2$ and the spin orientation signs 
$\sigma_1=\sigma_3=1$ and $\sigma_2=\sigma_4=-1$ according to Fig.\ \ref{fig:mnte}, 
we find the effective hopping $\tilde{t}=72.34$ meV
and the effective exchange $\tilde{J}=1.74$ meV. This leads to the MBS of the charge-transfer gap
\begin{align}
\label{eq:mnte}
\Gamma_{\rm CTG}(0)&=\frac{C_0}{2}\tilde{t}+\frac{C_1}{2}\tilde{J}+\frac{C_2}{2}\frac{\tilde{J}^2}{\tilde{t}} \nn \\
&\approx (95+37-2)\ {\rm meV}=130\ {\rm meV} \ ,
\end{align}
where the expansion coefficients for $S=2$ in Table \ref{tab:ec} are used. 
In Ref.\ \cite{Hafez-Torbati2021} we performed an extensive explicit analysis of the HKH model involving hopping and exchange 
interactions up to the fourth neighbor with the above parameter values for $\alpha$-MnTe.
Our results provided a very nice description of the experimental data \cite{Ferrer-Roca2000,Bossini2020} 
for the MBS in $\alpha$-MnTe. We obtained an MBS of the charge-transfer gap $\Gamma_{\rm CTG}(0) \approx 120$ meV. 
This indicates that Eq.\ \eqref{eq:mnte} has reproduced the result of the explicit analysis with only $8\%$ error.

We check further the results in Eq.\ \eqref{eq:mnte} by comparing the hopping and the exchange contributions 
with the results of the explicit calculations in Ref.\ \cite{Hafez-Torbati2021} individually. 
In Ref.\ \cite{Hafez-Torbati2021} we 
carried out a polynomial fit similar to Eq.\ \eqref{eq:mbs:exp}
and found that the MBS $120$ meV is consisted of a hopping contribution 
of about $76$ meV and an exchange contribution of about $43$ meV. Comparing with the results in Eq.\ \eqref{eq:mnte}
we identify an error of $25\%$ for the hopping contribution and an error of about $14\%$ for the exchange 
contribution, which still depicts overall a nice agreement. We stress that the results in Eq.\ \eqref{eq:mnte} 
are achieved by just some simple substitutions while the results in Ref.\ \cite{Hafez-Torbati2021} are obtained 
through extensive theoretical calculations. 

We would like to mention that if the Hubbard interaction changes from $U=4$ eV to $U=7$ eV 
and the Hund coupling from $J_{\rm H}=0.7$ eV to $J_{\rm H}=1.0$ eV the MBS in Eq.\ \eqref{eq:mnte} changes from 
$\Gamma_{\rm CTG}(0)\approx 120$ meV to $\Gamma_{\rm CTG}(0)\approx 140$ meV. This change is solely due 
to the hopping contribution as the exchange interactions are kept fixed. This shows that there is no 
significant dependence of the MBS on the Hubbard interaction and the Hund coupling. One only needs to have 
a rough estimate of these parameters; the exchange interactions are the essential ones.

\subsection{NiO}
\begin{figure}[t]
 \includegraphics[width=0.65\columnwidth]{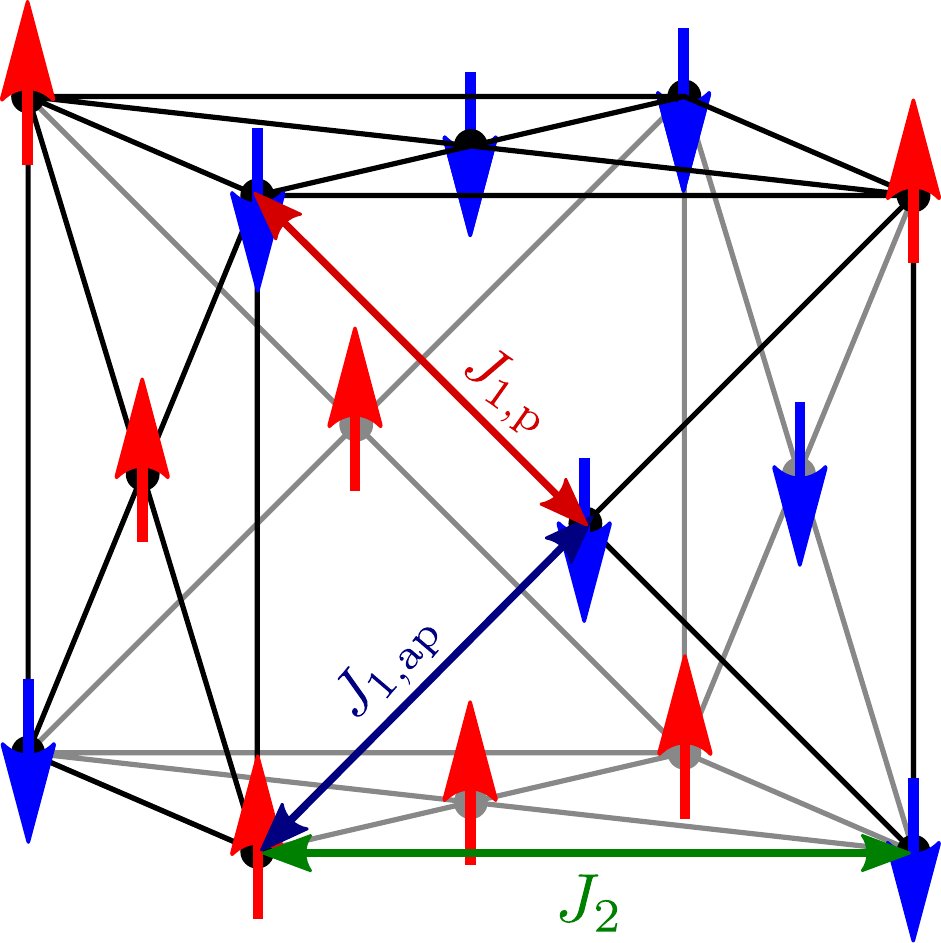}
 \caption{Magnetic structure of NiO. The second neighbor exchange interaction ($J_2$) is always between sites 
 with antiparallel spin orientation. The first neighbor exchange interaction can link sites with parallel ($J_{1,{\rm p}}$) or 
antiparallel ($J_{1,{\rm ap}}$) spin ordering.}
 \label{fig:nio}
\end{figure}

We proceed to the antiferromagnetic insulator NiO which has the N\'eel temperature $T_{\rm N}\approx 530$ K.
The magnetic Ni$^{+2}$ ions form ferromagnetic $(111)$ planes stacked antiferromagnetically on a face-centered 
cubic lattice \cite{Shull1951,Roth1958,Chatterji2009}. The magnetic structure of NiO is shown in Fig.\ \ref{fig:nio}.
Each magnetic ion is surrounded by twelve first neighbors having parallel alignment with six of them 
and antiparallel alignment with the other six. The alignment with the second neighbors is always antiparallel.
The onset of the magnetic ordering is found to be accompanied by a small rhombohedral distortion
which continues to zero temperature \cite{Slack1960,Bartel1971}. The lattice distortion is also expected 
to induce a band gap blue shift \cite{Ferrer-Roca2000,Bossini2020} which is described by an empirical 
Varshni function \cite{Varshni1967}.

Measurement of the magnon dispersion by inelastic neutron scattering suggest first neighbor ferromagnetic
exchange interactions $J_{1,{\rm p}}=-1.39$ meV and $J_{1,{\rm ap}}=-1.35$ meV and a second neighbor antiferromagnetic 
exchange interaction $J_{2}=19.01$ meV \cite{Hutchings1972}. The term $J_{1,{\rm p}}$ links sites  
with parallel spin ordering and the term $J_{1,{\rm ap}}$ links sites with antiparallel spin ordering, see Fig.\ \ref{fig:nio}. 
The small difference between $J_{1,{\rm p}}$ and $J_{1,{\rm ap}}$ originates from the lattice distortion. 
The antiferromagnetic interaction $J_2$ is much larger than the ferromagnetic 
interactions $J_{1,{\rm p}}$ and $J_{1,{\rm ap}}$. This suggests our approach to be applicable to NiO.

The Hubbard interaction and the Hund coupling in NiO were estimated from the local density approximation
combined with the DMFT as $U=8$ eV and $J_{\rm H}=1$ eV \cite{Ren2006}. The electron configuration $3d^8$
of the Ni$^{+2}$ ions implies that two orbitals need to be taken into account corresponding 
to the HKH model \eqref{eq:hkh} with the localized spin $S=1/2$. We set 
the first neighbor hopping $t_1=0$ due to the ferromagnetic exchange interaction and find the second neighbor
hopping $t_2=\sqrt{\Delta J_2}/2\approx 206.8$ meV. The effective exchange parameter in Eq.\ \eqref{eq:gj} and the 
effective hopping parameter in Eq.\ \eqref{eq:gt} can easily be obtained as $\tilde{J}=J_2+J_{1,{\rm ap}}-J_{1,{\rm p}}=19.05$ meV 
and $\tilde{t}=t_2 \approx 206.8$ meV.
Using the expansion coefficients provided in Table \ref{tab:ec} for $S=1/2$, we find the MBS of the 
charge-transfer gap in NiO
\begin{align}
\label{eq:nio}
\Gamma_{\rm CTG}(0)&=\frac{C_0}{2}\tilde{t}+\frac{C_1}{2}\tilde{J}+\frac{C_2}{2}\frac{\tilde{J}^2}{\tilde{t}} \nn \\
&\approx (120+269-29)\ {\rm meV}=360\ {\rm meV} \ ,
\end{align}
which is almost three times larger than the MBS we found for $\alpha$-MnTe in Eq.\ \eqref{eq:mnte}. 
This large MBS in NiO, despite the small localized spin, stems from the large effective exchange
interaction $\tilde{J}=19.05$ meV in contrast to the effective exchange interaction $\tilde{J}=1.74$ meV 
in $\alpha$-MnTe.

The optical absorption spectra of NiO has been the subject of extensive research in the past decades 
\cite{Newman1959,Powell1970,Huefner1984,Sawatzky1984,Tjernberg1996,Kuo2017} and 
a band gap of about $3.5$-$4.3$ eV is suggested depending 
on the method used to exctract the gap from the experimental data \cite{Huefner1992}. 
The temperature dependence of the band gap measured from the N\'eel temperature 
to the room temperature indicates an increases of about $350$ meV \cite{Jamal2019}.
If we assume based on our numerical results \cite{Hafez-Torbati2021} on the temperature dependences 
that below room temperature there is no significant further magnetic contribution 
to the blue shift of the band gap as the magnetization tends to saturate, the Varshni function fitted 
to the experimental data \cite{Terlemezoglu2021} captures the band gap blue shift due to the lattice 
distortion in NiO. Subtracting the lattice distortion contribution from the blue shift of the band gap 
$350$ meV 
one finds an MBS of about $300$ meV. This is in an overall nice agreement with our estimate $360$ meV 
in Eq.\ \eqref{eq:nio}.

\subsection{BiFeO$_3$}

\begin{figure}[t]
 \includegraphics[width=0.52\columnwidth]{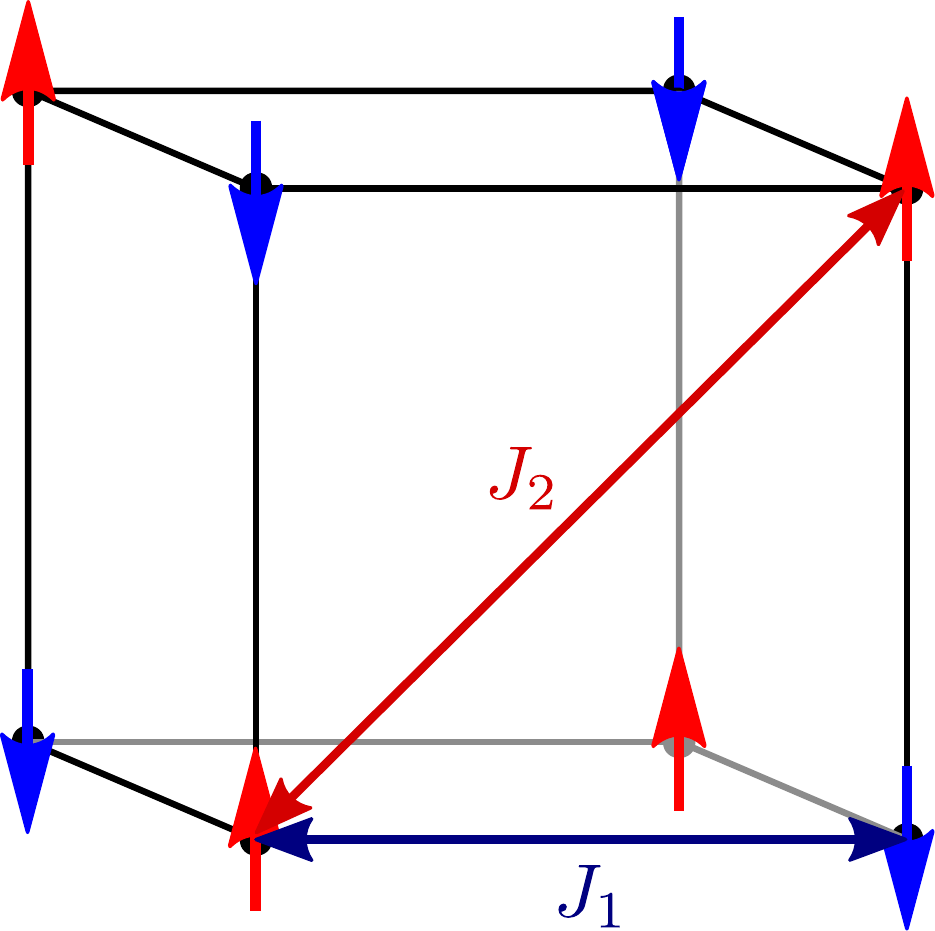}
 \caption{N\'eel antiferromagnetic order on a simple cubic lattice describing the magnetic structure
in BiFeO$_3$. The first neighbor exchange interaction $J_1$ links sites with antiparallel
spin ordering and the second neighbor exchange interaction $J_2$ links sites with parallel spin
ordering.}
 \label{fig:bfo}
\end{figure}

We now turn to the multiferroic semiconductor BiFeO$_3$ which has a band gap 
of about $2.7$ eV at the room temperature. 
The transition metal ions Fe$^{+3}$ with a half 
filled $3d$ shell develop a N\'eel antiferromagnetic order on a pseudo-cubic lattice structure below 
the N\'eel temperature $T_{\rm N}\approx 640$ K, see Fig.\ \ref{fig:bfo}. Each Fe$^{+3}$ magnetic moment is surrounded 
by six first neighbor Fe$^{+3}$ with an antiparallel magnetic moment alignment 
and twelve second neighbor Fe$^{+3}$ with a parallel magnetic moment 
alignment \cite{Kiselev1963,Blaauw1973,Jeong2012}.

The exchange interactions in BiFeO$_3$ are determined by the inelastic neutron scattering as 
$J_1=4.38$ meV for the first neighbor and the much weaker $J_2=0.15$ meV for the second neighbor \cite{Jeong2012}.
A Hubbard interaction $U=5$ eV and a Hund coupling $J_{\rm H}=1$ eV sound plausible from the 
different density functional theory analyses \cite{Clark2007,Neaton2005,Baettig2005}. 
These values result in the first neighbor hopping $t_1 \approx 99.3$ meV
and the second neighbor hopping $t_2 \approx 18.4$ meV. The effective hopping and the effective exchange interaction
can be found from Eqs. \eqref{eq:gt} and \eqref{eq:gj} as $\tilde{t} \approx 62.5$ meV and $\tilde{J}=4.08$ meV.
Using the expansion coefficients for $S=2$ in Table \ref{tab:ec} one obtains
\begin{align}
\label{eq:bfo}
\Gamma_{\rm CTG}(0)&=\frac{C_0}{2}\tilde{t}+\frac{C_1}{2}\tilde{J}+\frac{C_2}{2}\frac{\tilde{J}^2}{\tilde{t}} \nn \\
&\approx (83+86-12)\ {\rm meV} = 157\ {\rm meV}
\end{align}
for the MBS of the charge-transfer gap in BiFeO$_3$.

Despite extensive experimental research there are conflicting conclusions on the direct or the indirect nature of 
the band gap in BiFeO$_3$ and its temperature evolution. A direct band gap for BiFeO$_3$ is found in 
Refs. \cite{Kumar2008,Hauser2008,Ihlefeld2008,Li2010,Basu2008} which indicates a blue shift of about 100 meV upon 
reducing the temperature from $T_{\rm N} \approx 640$ K to zero \cite{Basu2008}.  
A much stronger change in the band gap is observed in Ref.\ \cite{Palai2008}. The change is
about $500$ meV between the N\'eel temperature and the room temperature  \cite{Palai2008}.   
A detailed Raman scattering study of the electronic band structure of BiFeO$_3$ associates the strong band gap 
blue shift observed in Ref.\ \cite{Palai2008} to the indirect nature of the band gap \cite{Weber2016}.
Band structure calculations based on the density functional theory also indicate an indirect band gap 
in BiFeO$_3$ \cite{Palai2008,Neaton2005,Clark2007}. 
More detailed investigations are needed to determine the temperature evolution of the band gap in 
BiFeO$_3$ and its microscopic origin, specifically including the effect of the antiferromagnetic ordering 
on the band gap blue shift. This will allow a comparison of the experimental data  
with our theoretical result in Eq.\ \eqref{eq:bfo} and can unveil the direct or the indirect 
nature of the band gap in this technologically attractive material.

\section{conclusion}
\label{sec:out}

We perform a systematic study of the effect of antiferromagnetic ordering on the single-particle 
spectral function in Mott insulators involving various number of orbitals. 
Our analysis relies on the DMFT of a HKH model. The model goes far beyond the 
low-energy Heisenberg model and permits investigation of charge excitations. 
The antiferromagnetic ordering is accompanied by a transfer of the spectral weight to lower energies and an MBS of the Mott gap. 
The MBS increases upon increasing the number of orbitals due to the double exchange and exchange mechanisms.

We provide an expansion for the MBS of the Mott gap in terms of the hopping and the exchange coupling in our prototype model.
We show how such an expansion can be generalized for application to realistic Mott or charge-transfer insulator materials.
This allows estimating the MBS in real materials in an extremely simple manner avoiding extensive theoretical calculations.
The approach is exemplarily applied to $\alpha$-MnTe, NiO, and BiFeO$_3$ as attractive compounds for antiferromagnetic 
spintronics application \cite{Baltz2018}. We find an MBS of about $130$ meV for $\alpha$-MnTe which is in an overall very good agreement 
with the previous theoretical calculations and experimental data. 
For NiO and BiFeO$_3$ we obtain an MBS of about $360$ and $157$ meV, respectively.
Our results for NiO also match the available experimental data. For BiFeO$_3$ there are currently controversial 
results in the literature on the direct or the indirect nature of the band gap and its temperature dependence. A systematic study 
of the MBS in BiFeO$_3$ and its comparison with our theoretical prediction can unveil the direct or the 
indirect nature of the band gap in this multiferroic compound.

We emphasize that the formula \eqref{eq:gmbs} is intended to establish a quick, ready-to-use estimate for the MBS.
Our results paves the path for identifying materials with a strong spin-charge coupling which is a prerequisite to go beyond the 
pure magnetic state in Refs.\ \cite{Bossini2016,Hashimoto2018,Bossini2019} and to realize a coupled spin-charge coherent 
dynamics at a femtosecond time scale.

\acknowledgments

This study was funded by the German Research Foundation (DFG) in the International 
Collaborative Research Centre TRR 160 (Project B8).

%

\appendix

\section{Spectral function for $n_b=4$}
\label{app:A}

\begin{figure}[t]
   \begin{center}
 \includegraphics[width=0.9\columnwidth,angle=-90]{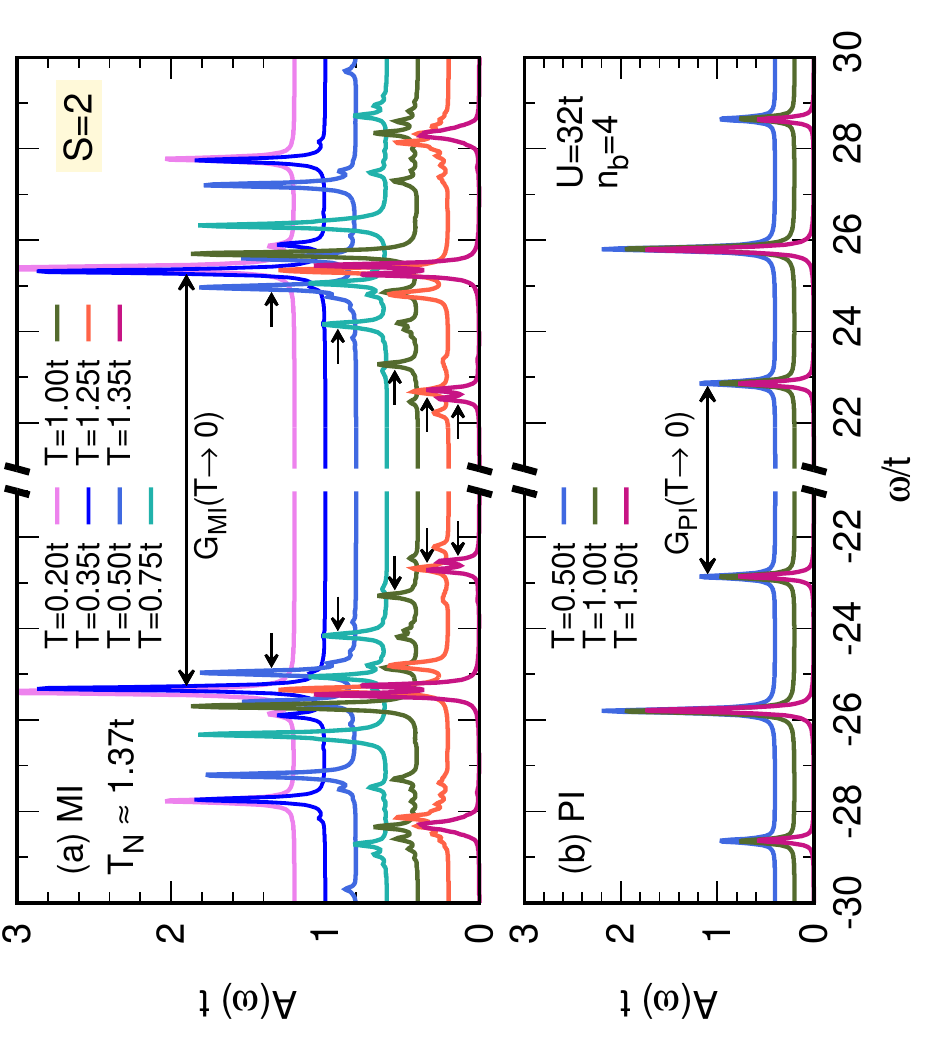}
   \caption{The spectral function $A(\omega)$ of the Hubbard-Kondo-Heisenberg model \eqref{eq:hkh} plotted vs frequency $\omega$ in the magnetic insulator (MI) phase (a) and 
   in the paramagnetic insulator (PI) phase (b) for the localized spin $S=2$ and the number of bath 
   sites $n_b=4$ in the ED impurity solver. The Hubbard interaction is fixed to $U=32t$ and the Hund coupling to $J_{\rm H}=0.3U/S$, which 
   result in the bare charge gap $\Delta=1.6U$. The spectral functions for the different temperatures $T$ are shifted 
   vertically for clarity. The N\'eel temperature is given by $T_{\rm N}\approx 1.37t$. The arrows specify the 
   peaks from which the Mott gap in the MI phase $G_{\rm MI}(T)$
   and in the PI phase $G_{\rm PI}(T)$ are read off.}
   \label{app:fig:sf:s2}
   \end{center}
\end{figure}

\begin{figure}[t]
   \begin{center}
 \includegraphics[width=0.9\columnwidth,angle=-90]{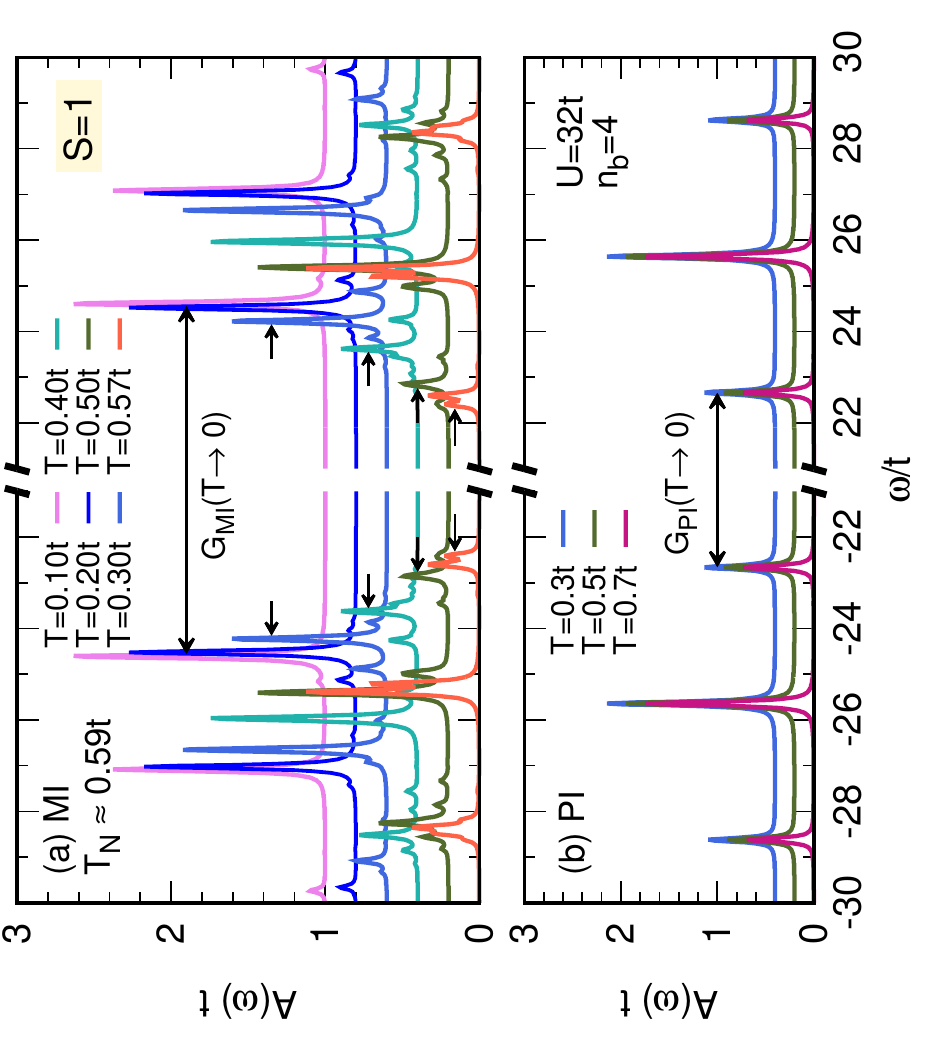}
   \caption{The same as Fig.\ \ref{app:fig:sf:s2} but for the localized spin $S=1$ with 
   the N\'eel temperature $T_{\rm N}\approx 0.59t$.}
   \label{app:fig:sf:s1}
   \end{center}
\end{figure}

We plot the spectral function $A(\omega)$ vs frequency at different temperatures $T$ in the MI phase  (a)
and in the PI phase (b) in Fig.\ \ref{app:fig:sf:s2} for the localized spin $S=2$ and in Fig.\ \ref{app:fig:sf:s1}
for the localized spin $S=1$. The results are shown for the bare charge gap $\Delta=U+2SJ_{\rm H}=1.6U$ for $U=32t$ 
and the number of bath sites $n_b=4$.
Similar to the results for the number of bath sites $n_b=6$ in Figs. \ref{fig:sf:s2} and \ref{fig:sf:s1} discussed 
in the main text, we find that the spectral function in the PI phase remains unchanged upon reducing the 
temperature, but in the MI phase there is a shift of the Mott gap specified by arrows to higher energies as
the temperature is reduced from the N\'eel temperature $T_{\rm N}$ to zero. We also find a similar magnetic 
redistribution of spectral weight as observed in the main text for $n_b=6$, i.e., there is a larger spectral weight near 
the Fermi energy in the MI phase at $T\to 0$ in contrast to the PI phase.

\end{document}